\documentclass[12pt,a4paper]{article}
\usepackage[utf8]{inputenc}
\usepackage[T1]{fontenc}
\usepackage[margin=2.5cm]{geometry}
\usepackage[numbered]{bookmark}
\usepackage{mathtools}
\usepackage{amssymb}
\usepackage{graphicx}
\usepackage[font=small,labelfont=bf]{caption}
\usepackage{authblk}
\usepackage{cite}
\usepackage{dsfont}
\usepackage{xcolor}
\usepackage[titles]{tocloft}

\numberwithin{equation}{section}

\newcommand{\eq}[1]{\begin{equation} #1 \end{equation}}
\newcommand{\al}[1]{\begin{align} #1 \end{align}}
\newcommand{\ml}[1]{\begin{multline} #1 \end{multline}}

\newcommand{\pa}{\partial}

\title{ More on Schrödinger holography }

\author[b]{A.~Golubtsova}
\author[a,b]{H.~Dimov}
\author[a]{I.~Iliev}
\author[a]{M.~Radomirov} 
\author[a,c]{R.~C.~Rashkov}
\author[a]{T.~Vetsov }

\affil[a]{\textit{Department of Physics, Sofia University,}\authorcr\textit{5 J. Bourchier Blvd., 1164 Sofia, Bulgaria}\vspace{5pt} \vspace{3pt}} 
\affil[b]{\textit{The Bogoliubov Laboratory of Theoretical Physics, JINR,}\authorcr\textit{141980 Dubna, Moscow region, Russia}\vspace{5pt} \vspace{3pt}} 
\affil[c]{\textit{Institute for Theoretical Physics, Vienna University of Technology,}\authorcr\textit{Wiedner Hauptstr. 8--10, 1040 Vienna, Austria}}

\date{}
	
\begin{document}
	 
\maketitle

\begin{abstract}
We find explicit solutions for giant magnons and spiky strings living on the Schrödinger $Schr_5 \times T^{1,1}$ and compute dispersion relations. The holographic dual field theory is conjectured to be  a non-local dipole-deformed CFT at strong coupling. We find that the dependence between conserved charges in the dispersion relations is transcendental, which is quite different from the most symmetric case of spherical internal space. Keeping the squashing parameter $b$ general allows us to take some limits and to compare our results to known cases.
\end{abstract}


\section{Introduction}

The subject of the gauge/string duality has made rapid strides in recent years and remains
the mainstream subject in string theory since its discovery.
The impressive developments have triggered a profound boost in our thinking not only about those theories but also about fundamental laws of Nature and particularly spacetime itself. 

Although widely accepted, there is no derivation of the AdS/CFT duality from first principles. This hinders wider applications of the gauge/string duality to other backgrounds and other gauge theories. Hence, understanding the origins of the gauge/string duality is a very important problem, for which the currently known mathematical apparatus is insufficient.

Having that, we are forced to focus on particular cases of  gauge/string duality.
The narrower topic of particular models in the AdS/CFT context is a subject of vigorous
activity worldwide. Indeed, finding explicit solutions provides important bits of information and techniques that will be needed for further developments.
As a result, the progress in this field has been very fast. The successful phenomenological applications of the holographic duality so far points rather to its general character, but certainly there is a long way to go until put the gauge/string duality is put into a full-fledged framework and/or its limitations understood. 

Addressing strong coupling phenomenon, recent developments provided information for conformal field theories living on the boundary of spacetime  allowing highest supersymmetry. Finding the spectrum of strings propagating on a generic curved background is extremely challenging task.
Integrability properties widely presented in AdS/CFT correspondence became a key tool to find exact solutions of both, string theory in AdS space and gauge theory on its boundary. The analysis of large variety of rotating strings, spinning strings, giant magnons, folded strings, spiky strings, pulsating strings and their gauge theory duals provided invaluable information about the theories on both sides of the duality. 
According to the hoplographic correspondence, the dispersion relations on the string(gravity) side map to the anomalous dimensions of the gauge theory operators. This knowledge is a key input for the two main programs - studying strong coupled field theories on the boundary and bulk reconstruction from boundary data. 

Important classes of string solutions allowing to go beyond the supergravity approximation of the holograhic duality are  giant magnons, spiky strings and pulsating strings. The dispersion relations on string theory side provides information of the anomalous dimensions of the operators at strong coupling on the gauge theory side.  The key tool in these studies is integrability \cite{Beisert:2010jr} allowing reliable information on both sides of duality.

Looking from more general point of view,  symmetries have always  been at the core of achievements in physics more than a century. The holographic duality is not an exception - it maps isometries of the bulk background to symmetries of the boundary gauge theory. Combined with integrability, this provides a powerful tool for extracting  reliable information on both sides of the duality. 
The majority of studies have been focused on the cases of high amount of symmetry while the less symmetric cases are still a big challenge. These however are rare cases in the real physical systems. That's why the problem of studying cases with less (super)symmetry is an important direction of investigations.

The first example of AdS/CFT correspondence is the duality between gravity in $AdS_5\times S^5$ and $\mathcal{N}=4$ supersymmetric Yang-Mills theory.  The isometries of the spherical part of the background turn out to be responsible for the maximal amount of supersymmetry on the gauge theory side. The deformations of $\mathcal{N}=4$ SYM reducing supersymmetry has been systematically studied by Leigh and Strassler \cite{Leigh:1995ep}. The first steps towards mapping these deformations to the deformations of the string background have been  initiated by Lunin and Maldacena in \cite{Lunin:2005jy}. The proposed simple mechanism advocated there enables to explicitly generate new background solutions suggesting, at the same time suggesting what holographic duals of these solutions could be. Shortly after that, Frolov formalized and further developed this solution generating technique that is now called a TsT transformation \cite{Frolov:2005ty}.
At practical level the procedure consists of T-duality, a shift with parameter $\beta$, followed by another T-duality applied to the isometry direction. Explicitly, let us assume that the background has a two torus submanifold parametriized  by ($\varphi_1 ,\varphi_2$). The TsT-transformation then consists of a T-duality along $\varphi_1$, followed by a shift $\varphi_2 \to \varphi_2 + \gamma\varphi_1$ in the T-dual background, and finally  by another T-duality along $\varphi_1$. 
A very short and incomplete  list of some applications of this method can be traced in \cite{Frolov:2005ty,frolovnew, Gursoy:2005cn, Chu:2006ae, Bobev:2006fg, Bobev:2007bm, Bobev:2005cz, Bykov:2008bj, Dimov:2009ut, Michalcik:2012mr, Frolov:2005dj} and references therein.


Going back to the role of symmetries in contemporary physics, one should note that nonrelativistic symmetries in theoretical physics have been always important. The Schr\"odinger equation, being the most fundamental element of quantum mechanics, has a large enough maximal symmetry group called the Schr\"odinger group. Due to its prominent role in theoretical physics, it deserves careful investigation in the context of the holographic correspondence. This line of investigations has been pioneered by Son \cite{Son:2008ye}, and by Balasubramanian and McGreevy \cite{Balasubramanian:2008dm,Adams:2008wt}.

The extension of AdS/CFT correspondence to non-relativistic field theories in D spatial dimensions, featuring an anisotropic scale law $(t,X^i)\to (\lambda^2t,\lambda X^i)$ in combination with special conformal transformations, points towards replacing AdS with spaces possessing   Schr\"odinger symmetry.  In the two papers \cite{Son:2008ye,Balasubramanian:2008dm} the geometry with Sch\"odinger isometries, dubbed also as Sch\"odinger spacetime, is proposed to be the gravitational background dual to non-relativistic field theories at strong coupling.

The non-relativistic limit of conformal symmetry is actually the Schr\"odinger algebra (for review see \cite{Dobrev:2013kha} and references therein). An incomplete list of some development in the context of AdS/CFT correspondence can be found in \cite{Duval:2008jg,Akhavan:2008ep,Bobev:2009zf}, as well as in \cite{Dobrev:2013kha}.

Recently there is a revival interest to non-relativistic version of AdS/CFT correspondence triggered by \cite{Guica:2017jmq}. By making use of Bethe ans\"atz techniques the authors investigated the integrability of holographic duality between $Schr_5\times S^5$ space and dipole field theories. It has been shown that Schr\"odinger holography matches the anomalous dimensions of certain gauge theory operators with calculation in BMN string limit. 
The established integrability and results of \cite{Guica:2017jmq} have motivated further study of the correspondence in non-relativistic cases, see for instance \cite{Ahn:2017bio,Georgiou:2017pvi,Georgiou:2018zkt,Dimov:2019koi,Georgiou:2019lqh,Georgiou:2020qnh,Ouyang:2017yko}. The BMN string spectrum has been considered in \cite{Ouyang:2017yko} while strings on Schr\"odinger pp-wave background \cite{Georgiou:2019lqh} have been complemented with the study of giant gravitons in the same background \cite{Georgiou:2020qnh}. The considerations of holographic three-point correlation function have been generalized to the case of Schr\"odinger/dipole CFT holography in \cite{Georgiou:2018zkt}.  Giant magnons and spiky strings in $Schr_5\times S^5$ background have been subject of investigation in \cite{Ahn:2017bio} and \cite{Georgiou:2017pvi}. Pulsating string holography in this background was also studied in \cite{Dimov:2019koi}.

In this paper we proceed with investigations of the Schr\"odinger holography, focusing on the giant magnon and spiky strings in $Schr_5\times T^{1,1}$. In the case of the Klebanov-Witten model \cite{Klebanov:1998hh}, i.e. $AdS_5\times T^{1,1}$/CFT holography, the supersymmetry is reduced and the dispersion relations are quite different \cite{Benvenuti:2008bd} from known ones in $AdS_5\times S^5$ case. This motivates us to thoroughly study its generalization to Schr\"odinger holography.

In the next section we quote the result for giant magnon and spiky string disppersion relations in the case of $Schr_5\times S^5$ background and set up the notations for $Schr_5\times T^{1,1}$ background. 

In the third section we focus on finding giant magnon and spiky string solutions. The fourth section is devoted to the dispersion relations of giant magnon and spiky strings in $Schr_5\times T^{1,1}$ background. The last section contains some concluding remarks and discussion on our findings.


\section{The setup}

In this Section we introduce our notations and formluae that will be needed,  mention some known results and make preparations for next Sections.\\

\textbf{General formulae and notations.} To fix the notations we start with the Polyakov string action  in general background.
\begin{equation}
S=-\frac{T}{2}\, \int d\tau d\sigma \left\{\sqrt{-h}\,h^{ab}\, \partial_a X^M \partial_b  X^N \,G_{MN} + \epsilon^{ab}\, \partial_a X^M \partial_b X^N \,B_{MN}\right\},
\end{equation}
where $a, b=0,1$ and $M,N=0,\dots, 9$.  We recall that the string tension T is related to the ’t Hooft coupling $\lambda\equiv g^2 _{YM}\,N\,\,$, as $\,\,T=\dfrac{\sqrt{\lambda}}{2\pi}$. The metric above is in conformal gauge $h^{ab}=\rm{diag}(-1,1)$ while for the antisymmetric tensor density we use the convention $\epsilon^{01}=-\epsilon^{10}=1$. In these notations the explicit form of the action is
\begin{equation}
S =-\frac{T}{2} \int \!d\tau d\sigma \left\lbrace  G_{MN} \left[ -\, \pa_\tau X^M \pa_\tau X^N + \pa_\sigma X^M \pa_\sigma X^N \right] + 2B_{MN} \,\pa_\tau X^M \pa_\sigma X^N \right\rbrace.
\label{action}
\end{equation}

It is accompanied with the standard Virasoro constraints 
\begin{align}
&\text{Vir}_1:\qquad G_{MN}\left(\partial_{\tau} X^M \partial_{\tau} X^N+\partial_{\sigma} X^M \partial_{\sigma} X^N\right)=0\,\label{vir1},\\
&\text{Vir}_2:\qquad G_{MN}\,\partial_{\tau} X^M \partial_{\sigma} X^N=0.\label{vir2}
\end{align}
As it is well known, according to AdS/CFT the dispersion relation on the string side corresponds to the anomalous dimension of the operators on gauge theory side. 
The conserved charges associated with the symmetries of the background are defined as follow
\begin{equation}
E=-\int\limits_{0}^{2\pi} \!d\sigma \,\frac{\pa \mathcal{L}}{\pa \left( \partial_{\tau} t\right)},
\qquad J_\phi = \int\limits_{0}^{2\pi} \!d\sigma \,\frac{\pa \mathcal{L}}{\pa \left( \pa_{\tau} \phi \right) },
\end{equation}
where $E$ is the energy and $\phi$ is an isometry direction.

%
For our purpose, namely to find solutions in the class of giant magnons or single spikes, it is convenient to choose another parametrization. We make the following change of worldsheet coordinates: $\tau\,\leftrightarrow\, \tau\,$ and $\sigma\,\rightarrow\, \xi = \alpha\sigma+\beta\tau\, ,\,\, \xi\in\,(-\infty,\, +\infty)$.\\

In the new parametrization the Polyakov action \eqref{action} takes the following convenient form
\begin{multline}\label{actionMag}
S=\int \!d\tau d\xi \,L=-\frac{T}{2\alpha} \int \!d\tau d\xi \left\lbrace  G_{MN} \left[ -\, \pa_\tau X^M \pa_\tau X^N - 2\beta\,\pa_\tau X^M \pa_\xi X^N \right. \right. \\
\left. \left. +\,(\alpha^2 -\beta^2)\,\pa_\xi X^M \pa_\xi X^N \right] 
+ 2\alpha \,B_{MN} \,\pa_\tau X^M \pa_\xi X^N \right\rbrace .
\end{multline}
The  Virasoro constraints, rewritten in these notations are
\begin{align}
&\text{Vir}_1:\quad G_{MN}\left( \pa_\tau X^M \pa_\tau X^N + 2\beta\,\pa_\tau X^M \pa_\xi X^N  + (\alpha^2 +\beta^2)\,\pa_\xi X^M \pa_\xi X^N\right) = 0 \label{vir1Mag},\\
&\text{Vir}_2:\quad  G_{MN} \left( \pa_\tau X^M \pa_\xi X^N + \beta\,\pa_\xi X^M \pa_\xi X^N\right) =0,\label{vir2Mag}
\end{align}
while the corresponding conserved charges become
\begin{equation}
E=-\int\limits_{-\infty}^{+\infty}\!d\xi \,\frac{\pa L}{\pa \left( \pa_\tau t\right) },
\qquad J_\phi = \int\limits_{-\infty}^{+\infty} \!d\xi \,\frac{\pa L}{\pa \left( \partial_\tau \phi \right) }.
\end{equation}
The equations of motion (EoM) in the new parametrization reads off  explicitly
\begin{align}
&\pa_\tau \left\lbrace   G_{KM}\, \pa_\tau X^M + \left(\beta\, G_{KM} -\alpha\,B_{KM} \right)    \pa_\xi X^M \right\rbrace \nonumber \\
&+ \pa_\xi \left\lbrace   -(\alpha^2 -\beta^2) \,G_{KM}\, \pa_\xi X^M + \left(\beta\, G_{KM} + \alpha\,B_{KM} \right) \pa_\tau X^M \right\rbrace \nonumber \\
&+\frac{1}{2}\frac{\pa G_{MN}}{\pa X^K} \left\lbrace -\,\pa_\tau X^M \pa_\tau X^N - 2\beta\,\pa_\tau X^M \pa_\xi X^N
+ (\alpha^2-\beta^2) \,\pa_\xi X^M \pa_\xi X^N \right\rbrace \nonumber \\
&+\alpha\,\frac{\pa B_{MN}}{\pa X^K} \,\pa_\tau X^M \pa_\xi X^N =0 .
\end{align}

Although we already set up the general formulae needed for obtaining dispersion relations, it is worth to pay a little attention of how to generate Schr\"oginger backgrounds. The procedure of TsT transformations outlined in the Introduction consists of several steps.  The starting point is $AdS_5$ space times Einstein manifold with compact isometry $\chi$
\eq{
ds^2=ds^2_{AdS_5}+ds^2_{M^5}, 
}
where
\eq{
ds^2_{AdS_5}=\ell^2\,\frac{2dx^+dx^-+dx^idx_i +dz^2}{z^2}, \qquad ds^2_{M^5}=ds^2_{C^4}+(d\chi+P)^2.
	\label{ads-l-c}
}

The procedure consists of T-duaity along $\chi$, then follows a shift $x^-\,\rightarrow\, x^-+\tilde{\mu}\tilde{\chi}$, where $\tilde{\chi}$ is T-dualized $\chi$. Finally T-dualization back on $\tilde{\chi}$ produces the metric
\eq{
	ds^2=\ell^2\left(- \frac{{\hat\mu}^2(dx^+)^2}{z^4} + \frac{2dx^+d\hat{x}^-+dx^idx_i+dz^2}{z^2} \right)+ ds^2_{\hat{M}^5} .
	\label{schro-metric}
}
Even though the initial background does not have B-field, the TsT transformed background acquires non-zero $B$-field:
\eq{
	\alpha' B_{(2)}=\frac{\ell^2{\hat\mu}\, dx^+}{z^2}\wedge (d\hat{\chi}+ P).
	\label{schro-B}
}
The summary of the results of these transformations is%
\eq{
	\frac{ds^2_{Schr_5}}{\ell^2}=-\left(\frac{{\hat\mu}^2}{Z^4}+1 \right)dT^2+
	\frac{2dT\,dV-\vec{X}^2dT^2+d\vec{X}^2+dZ^2}{Z^2},
	\label{metric-schro-global-a}
}
\eq{
	\alpha' B_{(2)}= \frac{\ell^2{\hat\mu}\, dT}{Z^2}\wedge (d\hat{\chi}+P), \qquad
	\hat\mu=\frac{\ell^2}{\alpha'}\tilde{\mu}=\sqrt{\lambda}\tilde{\mu}=\frac{\sqrt{\lambda}}{2\pi}L.
	\label{B-global-1-a}
}
It is easy to find the relation between original and dualized coordinates
\al{
	d\chi= d\hat{\chi} + \hat{\mu}\frac{dx^+}{z^2} \label{dual-psi},\qquad dx^-=d\hat{x}^- -\hat{\mu}\left(d\hat{\chi} +\hat{\mu} \frac{dx^+}{z^2} +P\right).
}
The dual coordinates satisfy periodic boundary conditions while the original coordinates satisfy twisted boundary conditions:
\begin{equation}
\label{eq_boundary_condition_1}
x^-(2\pi)-x^-(0)= LJ,
\end{equation}
\begin{equation}\label{eq_boundary_condition_2}
\chi(2\pi)-\chi(0)=2\pi m - LP_-.
\end{equation}


\subsection{The Schr\"odinger backgrounds $Schr_5 \times S^5$ and $Schr_5 \times T^{1,1}$}

In this subsection we provide explicit formulae for the metrics and  known results.\\

\textbf{The case of $Schr_5 \times S^5$ background.}
Below we give a very brief summary of the results from studies of giant magnon and spiky strings on $Schr_5 \times S^5$ background \cite{Ahn:2017bio,Georgiou:2017pvi}. These specific string solutions are of the form of arcs or spikes traveling along certain geodesics. 

We start with the metric of Schr\"odinger space $Schr_5$ in global coordinates (see Appendix \ref{appSchro})
\begin{equation}\label{metric Schr}
ds^2_{Schr_5}=-\left(1+\frac{{\mu}^2}{Z^4} + \frac{\vec{X}^2}{Z^2}\right)dT^2 +\frac{2dTdV+d\vec{X}^2+dZ^2}{Z^2}.
\end{equation}
The metric of the round $S^5$ is useful to write as a  Hopf fibration over the base $\mathbb{CP}^2$
\begin{equation}
ds^2_{S^5}=ds_{\mathbb{CP}^2}^2+(d\chi+P)^2.
\end{equation}
Here $P$ is a differential on the base
\begin{equation}
P=\frac{1}{2}\sin^2\mu\left(d\alpha+\cos\theta\,d\phi\right).
\end{equation}

To present the results in this case we stuck to the notations in \cite{Georgiou:2017pvi} where the considerations are restricted to $\vec{X}=0$ and $S^3\subset S^5$. In this case the relevant formulae are (in the units $\ell=\alpha'=1$)
\begin{align}
& ds^2= -\left(1+\frac{\mu^2}{Z^4}\right)dT^2+\frac{2dTdV+dZ^2}{Z^2}+\frac{1}{4}\left[d\theta^2+\sin^2\theta d\phi^2+(d\psi-\cos\theta d\phi)^2 \right], \\
& B=\frac{\mu}{2Z^2}dT\wedge(d\psi-\cos\theta d\phi).
\end{align}

The specific ans\"atz applied in that paper is
\al{
& T=\kappa\tau + T_y(y), \qquad V=\alpha\tau +V_y(y), \qquad Z=Z_0, \\
& \theta=\theta_y(y), \qquad \psi=\omega_\psi\tau + \Psi_y(y), \qquad \phi=\omega_\phi\tau + \Phi_y(y),
}
where $y=c\sigma-d\tau$ and $\kappa,\alpha,Z_0,\omega_\psi,\omega_\phi$ are constants.

Beside the four charges ($\mathcal{E},\mathcal{M},\mathcal{J}_\psi,\mathcal{J}_\phi$) resulting from integration of the momenta $p_t,p_V,p_\psi$ and $p_\phi$, it is useful to introduce the quantity $\Delta\varphi_1=(\Delta\psi+\Delta\phi)/2$.

After lengthy calculations, the dispersion relations for giant magnon strings are found to be
\eq{
\left(\sqrt{\mathcal{E}^2-\mu^2\mathcal{M}^2}-\mathcal{J}_1 \right)^2-\mathcal{J}_2^2=4\sin^2\theta\frac{\Delta\varphi_1}{2},
}
where $\mathcal{J}_{1/2}= \mathcal{J}_\psi\pm \mathcal{J}_\phi$.

Analogous calculations for the single spike strings give the dispersion relations
\eq{
\frac{1}{4}(\mathcal{J}_1^2-\mathcal{J}_2^2) =\sin^2\left[\frac{1}{2}(\mathcal{E}-\mu\mathcal{M}-\Delta \varphi_1) \right].
}\\



\textbf{The case of $Schr_5 \times T^{1,1}$ background.}

Let us set up the notation of the line element of  $Schr_5 \times T^{1,1}$ background
geometry (\cite{Guica:2017jmq}, \cite{vanTongeren:2015uha}, \cite{Gauntlett:2004yd}, \cite{Cvetic:2005ft})
\begin{equation}
ds^2_{Schr_5\times T^{1,1}}=ds^2_{Schr_5} + ds^2_{T^{1,1}},
\end{equation}
which will be written in global coordinates (see Appendix \ref{appSchro}).
The metric of $Schr_5$ in global coordinates is given in \eqref{metric Schr} while that for $T^{1,1}$ reads
\begin{equation}\label{metric T11}
ds^2_{T^{1,1}}=\frac{b}{4}\!\left[\sum\limits_{i=1}^2\left(d\theta_i^2+\sin^2\theta_id\phi_i^2\right)
+b\!\left(d\psi - \sum\limits_{i=1}^2\cos\theta_id\phi_i\right)^{\!\!2}\,\right]\!,
\end{equation}
where $0\leq\psi<4\pi,\ 0\leq\theta_i\leq\pi,\ 0\leq\phi_i<2\pi$, and $b=2/3$. We will keep the parameter $b$ because taking limit $b\to 1$ the metric of $S^5$ will be recovered and we can compare our results with those in \cite{Georgiou:2017pvi}.
The explicit matrix form of the above metrics (with ordering $(\theta_1,\,\theta_2,\,\phi_1,\,\phi_2,\,\psi )$) is
\begin{equation}
	\left(\hat{G}_{kh}^{T^{1,1}}\right)=\dfrac{b}{4}\left(\begin{matrix}
		1 & 0& 0&0&0\\
		0&1&0&0&0\\
		0&0&b\cos^2\theta_1+\sin^2\theta_1 & b\cos\theta_1\cos\theta_2 & -b\cos\theta_1\\
0&0& b\cos\theta_1\cos\theta_2 & b\cos^2\theta_2+\sin^2\theta_2 & -b\cos\theta_2\\
0&0& -b\cos\theta_1 & -b\cos\theta_2 & b
	\end{matrix}\right)\!.
\end{equation}

Let us remind that TsT transformations used to generate Schr\"odinger background produce a $B$-field which in our case has the form
\begin{equation}
B_{(2)} =
\dfrac{b\mu}{2 Z^2} \,dT\wedge \left( d\psi - \sum\limits_{i=1}^2\cos\theta_i \,d\phi_i \right) .
\end{equation}	
The B-field components are then
\begin{equation}
B_{T \phi_1}=-\,\dfrac{b\mu}{2Z^2}\,\cos\theta_1\,,  \qquad   B_{T\phi_2}=-\,\dfrac{b\mu}{2 Z^2}\,\cos\theta_2\,, \qquad B_{T\psi}=\dfrac{b\mu}{2Z^2}.
\end{equation}

Having the set up done, we are in a position to proceed with finding specific solutions. 
%
%
\subsection{The ans\"atz, Lagrangian and conserved charges}

To find solitary type string solutions we need to introduce a specific ans\"atz. As we mentioned above, these solutions represent solid string profiles traveling along geodesics. Thus, the ans\"atz we will use to find classical solutions (see for instance also \cite{Dimov:2007ey}, \cite{Benvenuti:2008bd}, \cite{Georgiou:2017pvi}) is
\begin{align}\label{ansatz}
&T =\kappa \tau + t(\xi),\,\,\kappa>0,\quad V=\omega_0 \tau+v(\xi),\quad\, Z=const\neq 0,\quad\, \vec{X}=\vec{0},\nonumber \\
&\theta_i=\theta_i (\xi), \qquad \phi_i=\omega_i\,\tau +\Phi_i(\xi), \quad i=1,2, \qquad
\psi=\omega_3\,\tau +\Psi(\xi).
\end{align}

Given above ans\"atz, explicit form of the string  Lagrangian ($ X^{\prime}=\partial_{\xi}X$) reads
\begin{multline}
L\, =-\, \frac{T}{2\alpha} \left\lbrace G_{TT} \left[   -\dot{T}^2 -2\beta\,\dot{T}T' +(\alpha^2-\beta^2) \,{T'}^2 \right] 
+2\,G_{TV} \left[ -\dot{T}\dot{V}-\beta\,(\dot{T}V'+\dot{V}T')\right. \right. \\  \left.
+(\alpha^2-\beta^2) \, T'V' \right] 
+(\alpha^2-\beta^2)\sum\limits_{i=1}^2 G_{\theta_i \theta_i} \,{{\theta_i}'}^2  
+\sum\limits_{i,j=1}^2 G_{\phi_i \phi_j} \left[ -\dot{\phi_i}\dot{\phi_j} - 2\beta \,\dot{\phi_i}  {\phi_j}'+(\alpha^2-\beta^2)\,  {\phi_i}'{\phi_j}'   \right]   \\
+2\sum\limits_{i=1}^2 G_{\phi_i \psi} \left[ -\dot{\phi_i}\dot{\psi}-\beta\,(\dot{\phi_i}\psi'+\dot{\psi}{\phi_i}') + (\alpha^2-\beta^2) \,{\phi_i}'\psi' \right]  
+G_{\psi \psi} \left[-\dot{\psi}^2 - 2\beta \,\dot{\psi}  \psi'+(\alpha^2-\beta^2)  \,{\psi'}^2 \right] \\
\left. + 2\alpha \sum\limits_{i=1}^2 B_{T\phi_i} \left[ \dot{T}{\phi_i}'-T'\dot{\phi_i} \right] + 2\alpha B_{T\psi} \left[ \dot{T}{\psi}'-T'\dot{\psi} \right]\,\right\rbrace.
\end{multline}

The background space has isometries, which correspond to the shifts in $\,T,\,V,\,\phi_1,\,\phi_2$ and $\psi$. The momentum densities associated to these isometries, are defined in  a standard way
\begin{equation}
\Pi_T =-\,\frac{\pa L}{\pa\dot{T} }\,,\qquad \Pi_{V}=\frac{\pa L}{\pa\dot{V} }\,,\qquad \Pi_{\phi_k}=\frac{\pa L}{\pa\dot{\phi_k} }\,,\qquad \Pi_{\psi}=\frac{\pa L}{\pa\dot{\psi} }.
\end{equation}
Then, the corresponding conserved charges are given by the integrals
\begin{align}
&E\,=\,-\int\limits_{-\infty}^{+\infty} \!d\xi \,\frac{\pa L}{\pa \dot{T}}\,, \qquad J_V\,=\,\int\limits_{-\infty}^{+\infty} \!d\xi \,\frac{\pa L}{\pa \dot{V}}\,, \nonumber  \\ 
& J_{\phi_k}\,=\,\int\limits_{-\infty}^{+\infty} \!d\xi \,\frac{\pa L}{\pa \dot{\phi_k} }\,, \qquad J_{\psi}\,=\,\int\limits_{-\infty}^{+\infty} \!d\xi \,\frac{\pa L}{\pa \dot{\psi} }\,.
\end{align}
\\

Next step is to obtain the explicit form of the momentum densities for the ansatz \eqref{ansatz}
\begin{align}
&\frac{2\alpha}{T}\,\Pi_T\,=\,-\,2\left( G_{TT}\,(\kappa +\beta\, t')+G_{TV}\,(\omega_0 +\beta\,v')-\alpha \sum\limits_{i=1}^2 B_{T\phi_i} \Phi_i'-\alpha\,B_{T\psi} \Psi' \right) , \\
&\frac{2\alpha}{T}\,\Pi_V\,=\,2 \,G_{TV}\,(\kappa + \beta\, t')\,,\\
&\frac{2\alpha}{T}\,\Pi_{\phi_k}\,=\,2\left( \sum\limits_{i=1}^2 G_{\phi_k \phi_i} \, (\,\omega_i +\beta\,\Phi_i')+ G_{\phi_k \psi}\,(\,\omega_3 +\beta\,\Psi') +\alpha\, B_{T\phi_k}\,t'\, \right) ,\,\, k=1,2\,,\\
&\frac{2\alpha}{T}\,\Pi_{\psi}\,=\,2 \left( \sum\limits_{i=1}^2 G_{\phi_i\psi} \,(\,\omega_i +\beta\,\Phi_i')+ G_{\psi\psi}\,(\,\omega_3 +\beta\,\Psi') +\alpha\, B_{T\psi}\,t'\right) .
\end{align}

These formulae will be used in what follows to obtain the dispersion relations.


\section{Giant magnon and spiky string soultions in $Schr_5 \times T^{1,1}$ background}

In the previous Section we setup the notations and the ans\"atz for giant magnon or single spike string configurations. In this Section we are going to find the relevant equations of motion for dynamical degrees of freedom and implement Virasoro constraints. We will analyze under which conditions the solutions of the equations of motion would be of giant magnon or spiky type. This analysis will fix and relate some otherwise arbitrary constants. We will find eventually giant magnon and spiky string solutions\footnote{For some details see Appendix \ref{calculations}}.

\subsection{Equations of motion and Virasoro constraints}

%
%
\textbf{Equations of motion.}
The nontrivial equations of motion coming from the variation of the action after imposing the ans\"atz \eqref{ansatz} are as follows.\\
---  for $t^{\prime}$ ---
\begin{equation}\label{eq_t}
(\alpha^2 - \beta^2) \,{t^{\prime}}(\xi)=A_V Z^2 +\beta \kappa.
\end{equation}
--- for $v^{\prime}$ ---
\begin{equation}\label{eq_v}
(\alpha^2 - \beta^2)\,v^{\prime}(\xi)\,= \frac{\alpha b\mu}{2} \left( \omega_3 - \sum\limits_{i=1}^2 \omega_i \,\cos\theta_i \right)
+ A_T Z^2  + (Z^4+\mu^2)A_V+\beta\omega_0\,.
\end{equation}
--- for $\Phi_k^{\prime}$ ---
\begin{equation}\label{eq_Phi_k}
(\alpha^2 - \beta^2)\,{\Phi_k}^{\prime}(\xi)\, =\,\frac{4}{b}\,\frac{(A_{\phi_k} +  A_{\psi}\,\cos{\theta_k})}{\sin^2 {\theta_k }} \,+\, \beta\omega_k  \,, \qquad k=1,2\,,
\end{equation}
--- for $\Psi^{\prime}$ ---
\begin{equation}\label{eq_Psi}
 (\alpha^2-\beta^2)\,\Psi'(\xi)=\frac{4}{b}\sum\limits_{i=1}^2\dfrac{A_{\phi_i}\,\cos\theta_i +A_{\psi}}{\sin^2 {\theta_i}} +\dfrac{4(1-2b)}{b^2}\,A_{\psi}-\frac{2\alpha\mu\kappa}{b Z^2} +\beta\omega_3\,.
\end{equation}
--- for $ Z $ ---
\begin{equation}\label{RelationZ-1}
A_V^2 \,Z^6 \,+\, A_V A_T \,Z^4 \,+\, \frac{\alpha\kappa}{b} \left(2\mu A_{\psi} - \alpha b\omega_0 \right)=0.
\end{equation}

--- for $\theta_k,\: k=1,2$ ---
\begin{multline}
(\alpha^2 - \beta^2)\,{\theta_1}^{\prime\prime}(\xi) +(1-b)\cos\theta_1\sin\theta_1 \left[ \,\omega_1^2 +2\beta\omega_1 \Phi_1' -(\alpha^2-\beta^2)\,  {\Phi_1'}^2    \right] \\
-b\sin\theta_1\cos\theta_2 \left[ \,\omega_1\omega_2 +\beta \,(\omega_1\Phi_2'+\omega_2  \Phi_1') -(\alpha^2-\beta^2)\,  \Phi_1' \Phi_2'   \right] \\
+b\sin\theta_1 \left[ \,\omega_1\omega_3 +\beta \,(\omega_1 \Psi'+\omega_3  \Phi_1') -(\alpha^2-\beta^2)\,  \Phi_1' \Psi' \right] \\
-\frac{2\alpha\mu}{Z^2} \sin\theta_1 \left[  \kappa\Phi_1'-t'\omega_1 \right] =0\, ,
\end{multline}
and
\begin{multline}
(\alpha^2-\beta^2)\,{\theta_2}^{\prime\prime}(\xi) +(1-b)\cos\theta_2\sin\theta_2 \left[ \,\omega_2^2 + 2\beta\omega_2 \,\Phi_2' -(\alpha^2-\beta^2)\, {\Phi_2'}^2    \right]\\
-b\sin\theta_2\cos\theta_1 \left[ \,\omega_1\omega_2 +\beta \,(\omega_1  \Phi_2'+\omega_2  \Phi_1') -(\alpha^2-\beta^2)\,  \Phi_1'\Phi_2' \right] \\
+b\sin\theta_2 \left[ \,\omega_2\omega_3 +\beta \,(\omega_2  \Psi'+\omega_3  \Phi_2') -(\alpha^2-\beta^2)\,  \Phi_2'\Psi' \right] \\
-\frac{2\alpha\mu}{Z^2} \sin\theta_2 \left[  \kappa{\Phi_2}^{\prime}-t^{\prime}\omega_2 \right] =0\, .
\end{multline}

These equations are accompanied with a relation between the constants coming from the equation for Z (see \eqref{EqZ})
\begin{equation}\label{RelationZ}
A_V^2 \,Z^6 \,+\, A_V A_T \,Z^4 \,+\, \frac{\alpha\kappa}{b} \left(2\mu A_{\psi} - \alpha b\omega_0 \right)=0.
\end{equation}


\textbf{Satisfying Virasoro constraints.}
Analysis of the equations of motion shows that the triple $\,(\,\theta_2 = \frac{\pi}{2},\,\,\omega_2 =0,\,\, A_{\phi_2}=0\,)$ is a solution to the equations of motion. Hence, the submanifold $\mathcal{M}$ defined by fixing $(\theta_2 ,\,\phi_2)=const$ and  \eqref{ansatz} is a consistent subsector of classical string theory on $\,\,Schr_5 \times T^{1,1}\,\,$. Thus, we can consider the dynamics of our string configuration on this submanifold and use from now on the notations $ \theta_1(\xi)\equiv \theta(\xi),\Phi_1(\xi)\,\equiv\,\Phi(\xi)$ to describe it.

The requirement global time to be proportional to the worldsheet one leads to a restriction on $t^{\prime}$, namely equation \eqref{eq_t} has to vanish, $t^{\prime}\,=\,0$. This fixes the constant $A_V$
\begin{equation}
A_V= -\,\frac{\beta \kappa}{Z^2}.
\end{equation}

Fixing $A_V$ modifies equations of motion for $ v(\xi),\Phi(\xi), \Psi(\xi)$ and $Z$ on the submanifold  $\mathcal{M}$. They become\\
---  for $v'$ ---
\begin{equation}\label{EoM_v}
(\alpha^2-\beta^2) \,v'(\xi)=\frac{\alpha b\mu}{2} \left( \omega_3 - \omega_1 \cos\theta \right)
+ \left[ A_T - \left( 1+\frac{\mu^2}{Z^4}\right) \beta\kappa \right]\! Z^2 + \beta\omega_0 \,,
\end{equation}
--- for $\Phi'$ ---
\begin{equation}\label{EoM_Phi}
(\alpha^2-\beta^2) \,\Phi'(\xi) =\frac{4}{b}\,\frac{(A_\phi +  A_\psi \cos{\theta})}{\sin^2 \theta} + \beta\omega_1 \,,
\end{equation}
--- for $\Psi'$ ---
\begin{equation}\label{EoM_Psi}
(\alpha^2-\beta^2) \,\Psi'(\xi) =\frac{4}{b} \,\dfrac{(A_\phi \cos\theta +A_\psi)}{\sin^2\theta} +\dfrac{4(1-b)}{b^2}A_\psi -\frac{2\alpha\mu\kappa}{bZ^2} +\beta\omega_3\,,
\end{equation}
--- for $ Z $ ---
\begin{equation}\label{EoM_Z}
\beta \left(\beta\kappa -A_T \right) Z^2 + \frac{\alpha}{b} \left(2\mu A_\psi -\alpha b \omega_0 \right)=0.
\end{equation}

We will often use the combination of \eqref{EoM_Psi} and \eqref{EoM_Phi} of the form
\begin{equation}\label{relation}
(\alpha^2-\beta^2) \left( \Phi'\cos\theta - \Psi'\right) + \beta \left( \omega_3 -\omega_1 \cos\theta \right) =\frac{4}{b^2} \left( \frac{\alpha b\mu\kappa}{2Z^2} -A_\psi\right).
\end{equation}
%

%
Now we are in a position to analyze the Virasoro constraints.
It turns out that it is more convenient instead of \eqref{vir1Mag} and \eqref{vir2Mag} to consider their linear combinations
\begin{align}
& Vir\,1: \quad G_{MN}\left( \pa_\tau X^M \pa_\tau X^N + (\alpha^2-\beta^2)\,\pa_\xi X^M \pa_\xi X^N\right)=0\,, \label{Vir1}\\
& Vir\,2: \quad  G_{MN}\left( \beta\,\pa_\tau X^M \pa_\tau X^N - (\alpha^2-\beta^2)\,\pa_\tau X^M \pa_\xi X^N\right) =0.\label{Vir2}
\end{align}
Then, using the equations for $ v,\, \Phi ,\, \Psi$ the constraint \eqref{Vir2} becomes:
\begin{equation}
\left( \kappa A_T +\omega_1 A_\phi + \omega_3 A_\psi  \right)Z^2 -\beta\kappa\omega_0=0\,,
\end{equation}
or
\begin{equation}\label{VirMix}
\omega_1 A_\phi + \omega_3 A_\psi = \left( \frac{\beta\omega_0}{Z^2} -A_T \right) \kappa \,.
\end{equation}
\\
Substituting the expressions for equations \eqref{EoM_v}, \eqref{EoM_Phi}, \eqref{EoM_Psi} and \eqref{relation} into the first Virasoro constraint \eqref{Vir1}, we obtain the following equation for $\theta(\xi)$
\begin{multline}\label{eq-theta}
(\alpha^2-\beta^2)^2\,  {\theta\,'\,}^2 = -\sin^2\theta \left\lbrace (\alpha^2-\beta^2)\,\omega_1^2 +\left[ \frac{4}{b}\,\frac{(A_\phi +A_\psi \cos\theta)}{\sin^2\theta} +\beta\omega_1 \right]^2   \right\rbrace \\
-(\alpha^2-\beta^2)\,b \,\left(\omega_3-\omega_1 \cos\theta \right)^2 - b\left[ \beta \left(\omega_3-\omega_1 \cos\theta \right) + \frac{4}{b^2} \left( A_\psi- \frac{\alpha b\mu\kappa}{2Z^2} \right) \right]^2 \\
-(\alpha^2-\beta^2)\,\frac{4}{b} \left( G_{TT}\,\kappa^2 +\frac{2\omega_0 \kappa}{Z^2} \right) .
\end{multline}
Note that on the submanifold $\mathcal{M}$ the $ G_{TT}$ component of target space metric is $G_{TT}= -\,|G_{TT}|=-\left( 1+ \frac{\mu^2}{Z^4} \right)$.

Concluding this subsection we remark that the only step we have to make is to  determine values of the constants corresponding to giant magnon or single spike string configurations.

%
\subsection{Turning points, boundary conditions and string solutions}
The semi-classical magnon/spike string solutions are described by open strings moving in a subspace of a some manifold \cite{Hofman:2006xt}.

The ans\"atz \eqref{ansatz} converts the equations of notion into those of effective point particle.
Thus, string configurations we analyze are reaching $\theta \,=\,\pi$ and it serves as turning points, namely  ${\theta}^{\,\prime}\,=\,0$  and $v^{\,\prime}\,=\,0$ at $\theta\,=\,\pi\,$.

A quick look at the right hand sides of \eqref{EoM_Phi}, \eqref{EoM_Psi} and \eqref{eq-theta} shows that their behavior at $\theta \,=\,\pi$ is problematic. However, requiring regularity one can use the expansion of dangerous terms around $\pi$ to fix some constants. For example, expanding the first term on the right hand side of \eqref{eq-theta}  around $\pi$,  $\frac{\left(A^2_\phi +A^2_\psi +2A_\phi A_\psi \cos\theta\right) }{\sin^2\theta} $,  one finds $\frac{A^2_\phi+A^2_\psi-2A_\phi A_\psi}{(\theta\,-\,\pi)^2} + \frac{1}{2}( A^2_\phi +A^2_\psi+ A_\phi A_\psi) + O( (\theta\,-\,\pi)^2 )$. Analyzing all dangerous terms in \eqref{EoM_Phi}, \eqref{EoM_Psi} and \eqref{eq-theta}, we observe that the requirement for regularity of ${\Phi}^{\prime}$, ${\Psi}^{\prime}$ and ${\theta}^{\,\prime}$ at $\theta\,=\,\pi$ leads to
\begin{equation}\label{requirement finiteness A}
A_\psi = A_\phi \equiv A \,.
\end{equation}

Let us discuss how the choice \eqref{requirement finiteness A} reflects on the equations we are going to solve.
Setting $\theta \,=\,\pi$ as a turning point, we expand the right hand side of \eqref{eq-theta}  around $\theta \,=\,\pi$. Taking into account \eqref{requirement finiteness A}, we find 
\begin{multline}\label{eq-theta-A}
(\alpha^2-\beta^2)^2\,  {\theta\,'\,}^2 = -\sin^2\theta \left\lbrace (\alpha^2-\beta^2) \,\omega_1^2 + \left[  \frac{4A}{b}\,\frac{(1+\cos\theta)}{\sin^2\theta} +\beta\omega_1 \right]^2 \right\rbrace \\
-(\alpha^2-\beta^2)\,b \,\left(\omega_3-\omega_1 \cos\theta \right)^2 - b \left[  \beta \left(\omega_3-\omega_1 \cos\theta \right) + \frac{4}{b^2} \left( A- \frac{\alpha b\mu\kappa}{2Z^2} \right) \right]^2 \\
-(\alpha^2-\beta^2)\,\frac{4}{b} \left(-|G_{TT}|\kappa^2 +\frac{2\omega_0 \kappa}{Z^2} \right) .
\end{multline}
Thus, the turning point condition provides to the following relations between the constants
\begin{equation}\label{turn-point-theta}
\left[ \beta\,(\omega_1+\omega_3) + \frac{4}{b^2} \left(A -\frac{\alpha b\mu}{2Z^2}\kappa \right) \right]^2 
=(\alpha^2-\beta^2) \left[\frac{4}{b^2} \left(  |G_{TT}|\kappa^2 -\frac{2\omega_0 }{Z^2} \kappa \right) - (\omega_1+\omega_3)^2 \right] . 
\end{equation}

The requirement $v'=0$ for equation \eqref{EoM_v} at the turning point $\theta=\pi$ leads to fixation of $A_T$ through $Z$: 
\begin{equation}
Z^2 \left(A_T -|G_{TT}|\beta\kappa \right) +\beta\omega_0 +\frac{\alpha b\mu}{2}\, (\omega_1+\omega_3)=0\,,
\end{equation}
or
\begin{equation}\label{turning-v}
A_T= -\,\frac{\alpha b\mu}{2Z^2} \,(\omega_1+\omega_3) +|G_{TT}|\beta\kappa - \frac{\beta\omega_0}{Z^2}\,.
\end{equation}

As a result of imposing conditions \eqref{requirement finiteness A}, \eqref{turning-v} the behavior of right hand sides of \eqref{EoM_v}, \eqref{EoM_Phi}, \eqref{EoM_Psi} and \eqref{eq-theta-A} in the vicinity of $\theta=\pi$  become  $O((\theta-\pi)^2)$ .  Actually the conditions \eqref{requirement finiteness A}, \eqref{turning-v} greatly simplify the dynamical equations we have to solve.  Let us summarize the simplifications of these equations 
\\
--- for $v'$ ---
\begin{equation}\label{EoM_v A}
(\alpha^2-\beta^2)\,v'(\xi)= -\,\frac{\alpha b\mu}{2}\, \omega_1\,(1+\cos\theta),
\end{equation}
--- for $\Phi'$ ---
\begin{equation}\label{EoM_Phi A}
(\alpha^2-\beta^2)\,\Phi'(\xi) =\frac{4A}{b}\,\frac{(1 + \cos\theta)}{\sin^2\theta} + \beta\omega_1 \,,
\end{equation}
--- for $\Psi'$ ---
\begin{equation}\label{EoM_Psi A}
(\alpha^2-\beta^2)\,\Psi'(\xi) =\frac{4A}{b}\,\dfrac{(\cos\theta +1)}{\sin^2\theta} +\dfrac{4(1-b)}{b^2}A -\frac{2\alpha\mu\kappa}{bZ^2} +\beta\omega_3\,.
\end{equation}
Looking at the above equations one can make an immediate conclusion. It is quite clear that the first thing to do is to solve the equation \eqref{eq-theta-A} for  $\theta(\xi)$.  Given its solution, one can integrate \eqref{EoM_v A}, \eqref{EoM_Phi A} and \eqref{EoM_Psi A} to find $v(\xi)$, $\Phi(\xi)$ and $\Psi(\xi)$ respectively.

We proceed with solving \eqref{eq-theta-A}. 
To make this, it is convenient to introduce a new function $u(\xi)$ 
\begin{equation}\label{u}
u(\xi) =\cos^2 \frac{\theta (\xi)}{2} \,, \qquad u(\xi) \in [0,1]\,.
\end{equation}
This choice converts the equation for $\theta$ into
\begin{equation}
{u(\xi)'}^2 =a_4u^4+a_3u^3+a_2u^2+a_1u+a_0\equiv P_4(u).
\end{equation}
By virtue of conditions \eqref{requirement finiteness A} and \eqref{turn-point-theta} $a_{0}=a_{1}=0$ and the differential equation for $\theta(\xi)$ \eqref{eq-theta-A} in terms of $u(\xi)$ becomes 
\begin{equation}\label{EqU}
u'\,^2 =u^2 \left( a_4\,u^2 + a_3\,u + a_2 \right)\,\equiv\, P_4(u)\geq 0.
\end{equation}
Here, the explicit expressions for coefficients of the polynomial $P_4(u)$ in terms of constants of the ans\"atz are 
\begin{equation}\label{a4}
a_4 =-\, \frac{4(1-b) \,\alpha^2\,\omega_1 ^2}{\left(\alpha^2-\beta^2 \right) ^2} < 0\,,
\end{equation}
\begin{equation}\label{a3}
a_3 =\frac{ 4(1-b)\,\alpha^2\,\omega_1^2 +4\alpha^2\,\omega_1^2 - 4b\alpha^2\,\omega_1 (\omega_1+\omega_3) + \dfrac{8\alpha\beta\mu\kappa\omega_1}{Z^2}}{\left(\alpha^2-\beta^2 \right) ^2}   \,,
\end{equation}
\begin{equation}\label{a2}
a_2 =\frac{-\,4\alpha^2\,\omega_1^2 + 4b\alpha^2\,\omega_1 (\omega_1+\omega_3) -\dfrac{8\alpha\beta\mu\kappa \omega_1}{Z^2}-\dfrac{16A^2}{b^2}}{\left(\alpha^2-\beta^2 \right) ^2}  \,.
\end{equation}
%
A direct inspection shows that the following equality holds
\begin{equation}\label{coefficient sum}
a_4 +a_3 +a_2 =-\,\frac{16\,A^2}{b^2 \left(\alpha^2-\beta^2 \right)^2}\,.
\end{equation}
%
%
%
%
%
\\
The equation \eqref{EqU}  could be interpreted as  energy conservation of the dynamics on the intersection of the submanifold $\mathcal{M}$ and $T^{1,1}$ driven by the quartic potential $U(u)\equiv -P_4(u)$. Notice that in our case $a_{4}\leq 0$ (being $0 <b\leq 1$) and $u(\xi) \in[0,1]$.
Of course, the validity of equation \eqref{EqU} describing the string configuration imposes the condition $P_{4}(u) \geq 0,$ where the zero of $U(u)=0$ is realized at the turning points.
%
\\

Completing the analysis of the turning point issues, we note that due to \eqref{turning-v}, the algebraic equation for constant $Z$ \eqref{EoM_Z} becomes
\begin{equation}\label{EoM_Z A}
Z^2 \left[ 2(\beta^2-\alpha^2)\,\omega_0 +\frac{4\alpha\mu}{b} A  +\alpha\beta b\mu \,(\omega_1+\omega_3) \right] =2\beta^2\mu^2\kappa \,.
\end{equation}
\\
Finally, substituting the conditions \eqref{requirement finiteness A} and \eqref{turning-v} into second Virasoro constraint \eqref{VirMix}, we obtain the following algebraic relation 
\begin{equation}\label{Vir-turning}
(\omega_1+\omega_3) \left(A- \frac{\alpha b\mu}{2Z^2} \kappa \right) =\beta \left(\frac{2\omega_0}{Z^2}\kappa -|G_{TT}|\kappa^2 \right) \!.
\end{equation}
%
Imposing the turning point conditions \eqref{requirement finiteness A} and \eqref{turning-v} results into a system of three algebraic equations \eqref{turn-point-theta}, \eqref{Vir-turning} and \eqref{EoM_Z A} for the constants $A$, $Z$ and $\kappa$. 

Next step is to use the system of equations \eqref{turn-point-theta} and \eqref{Vir-turning} 
to eliminate $\left( \frac{2\omega_0}{Z^2}\kappa -|G_{TT}|\kappa^2 \right) $ in the first equation. The result is a quadratic algebraic equation for $\frac{4}{b^2}\left( A -\frac{\alpha b\mu}{2Z^2}\kappa \right)$ involving parameters $\omega_1,\omega_3$. The solutions of this equation determine whether the classical string configurations are of giant magnon or single spike type. The two types solutions are classified as follows
\begin{align}
&\frac{4}{b^2} \left( A -\frac{b\alpha \mu}{2Z^2}\kappa \right)= -\,\beta\,(\omega_1+\omega_3)  \qquad \text{giant magnons}\,,\label{magnon}\\
&\frac{4}{b^2} \left( A -\frac{b\alpha \mu}{2Z^2}\kappa \right)= -\,\frac{\alpha^2}{\beta} 
\,(\omega_1+\omega_3)  \qquad \text{single spike strings}\,.\label{spike}
\end{align}
It should be mentioned that taking the limit of zero deformation $\mu\rightarrow 0$ we obtain corresponding expressions for $A_\phi$ in the case of $AdS_5\times T^{1,1}$ \cite{Benvenuti:2008bd}.
The next step is to solve the two algebraic systems \eqref{magnon}, \eqref{Vir-turning} and \eqref{EoM_Z A} and \eqref{spike}, \eqref{Vir-turning} and \eqref{EoM_Z A} for the giant magnons and the single spike respectively, concerning constants 
\begin{align}
&A_m =\frac{\alpha b}{2} \left[ \frac{\omega_0}{\mu} -\frac{\beta b}{2\alpha} \, (\omega_1+\omega_3) \right]  \qquad \text{giant magnons}\,,\label{A magnon}\\
&A_s =\frac{\alpha b}{2}\,\frac{\omega_0}{\mu} \qquad\qquad \text{single spike strings}.
\label{A spike}
\end{align}
By these choices the parameters $\kappa^2$ and $Z$ are also fixed. For $\kappa$ one finds
\begin{align}
&\kappa_m^2 = \frac{\omega_0^2}{\mu^2} + \frac{b^2}{4}\, (\omega_1+\omega_3)^2  \qquad \text{giant magnons}\,,\label{kappa magnon}\\
&\kappa_s^2 = \frac{\omega_0^2}{\mu^2} \qquad\qquad \text{single spike strings}\, ,\label{kappa spike}
\end{align}
and for $Z^2$
\begin{align}
&Z_m^2 = \frac{\mu^2\kappa_m}{\omega_0}  \qquad\qquad \text{giant magnons}\,,\label{Z magnon}\\
&Z_s^2 = \frac{2\beta\mu^2 \kappa_s}{2\beta\omega_0 + b\mu\alpha\,(\omega_1+\omega_3)} \qquad \text{single spike strings}\,. \label{Z spike}
\end{align}

The above analysis exhausts the conditions that can be imposed on the parameters to determine the giant magnon and single spike string solutions. 
%

In order to determine the dispersion relation for the string configurations giant magnons and single spike, we need to solve \eqref{EqU}. The polynomial $P_4(u)\geq 0,$ has a double zero at $u=0$ and two others for the quadratic factor.  Supplemented with the requirements $\theta=\pi,\, (u=0)$ to be a turning point and the boundary conditions, one concludes that the u-equation has the following structure
\begin{equation}\label{eq u roots}
u'\,^2 = u^2\,(\,a_{4} \,u^2 +a_{3}\, u +a_{2}\,) = |a_{4}|\,u^2\,( r_{+} -u) \,(u +|r_{-}|)\,,
\end{equation}
where the roots $r_{-}$ and $r_{+}$ have been ordered as follows
\begin{equation}\label{order roots}
r_{-} \leq 0 \leq u(\xi) \leq r_{+} \leq 1\,.
\end{equation}
The solution of the above equation is simple and given by the expression
\begin{equation}\label{solution cosh^2}
u(\xi) = \frac{r_+ \,|r_-|}{(r_+ +|r_-|)\,\cosh^2 \left(   \dfrac{\sqrt{|a_4|\, r_{+} \,|r_-|}}{2}\,\xi \right) -\,r_+}\,,
\end{equation}
or using that $2\cosh^2x =1+\cosh(2x)$ one gets
\begin{equation}\label{solution cosh}
u(\xi) = \dfrac{\frac{2\,r_+ \,|r_-|}{(r_+\,+\,|r_-|)}}{\cosh\left( \sqrt{|a_4|\,r_{+} \,|r_-|} \,\,\xi \right) -\, \dfrac{r_+ -|r_-|}{r_+ +|r_-|}}\,.
\end{equation}

The next issue is how to relate the roots of polynomial $P_4$ entering the solution to fixed above constants.
First of all we use the Vieta's formulas to express the roots in terms of $a_i,\:i=2,3,4$
\begin{equation}\label{Viet}
r_+ - |r_-| =\frac{a_{3}}{|a_{4}|}\,,\qquad r_+ \,|r_-| =\frac{a_{2}}{|a_{4}|} >0\,,\, \,i.e. \qquad a_2 >0\,.
\end{equation}
Next, we have to obtain the relation between constant $A$ and the coefficients $a_i$.
It is a simple exercise using \eqref{coefficient sum}, combined with Vieta's formulas, to obtain an explicit expression for $A$ in terms of the roots $r_-$ and $r_+$
\begin{equation}\label{A^2 roots}
\frac{16 A^2}{b^2\, (\alpha^2-\beta^2)^2} = -\,a_{4} \left( \frac{a_2}{a_4} + \frac{a_3}{a_4} +1 \right) = |a_4|\,(1 - r_{+}) \,(1+ |r_{-}|) \geq 0 \,,
\end{equation}
or
\begin{equation}\label{A roots}
\frac{2A}{b} = \pm \, |\alpha|\,|\omega_1|\,\sqrt{(1-b)\,|1- r_{+}|\,(1+ |r_{-}|) }\,.
\end{equation}
\\
Finally, putting everything together one can rewrite the solution \eqref{solution cosh} in the following form
\begin{equation}\label{solution}
u(\xi) = \frac{\dfrac{ 2a_2}{\sqrt{a_3^2 -4a_4 a_2}}}{\cosh\left( \sqrt{a_2} \,\,\xi\right) -\, \dfrac{a_3}{\sqrt{a_3^2 -4a_4 a_2}}}\,.
\end{equation}
We remark that $a_2 >0$ and $\frac{|a_3|}{\sqrt{a_3^2 -4a_4 a_2}} <1$. The above formula is a well defined solitary wave solution (see for instance \cite{Benvenuti:2008bd}).

\section{Dispersion relations}

In this Section we obtain the dispersion relations for the case of giant magnon and single pike string solutions. In the previous sections we have imposed several conditions, have obtained some relations between parameters and have found the general solutions of solitary wave type. First thing to do now is to apply all the findings so far to the momentum densities and conserved charges. After that we will find the dispersion relations for the two types string configurations.
%
%
%

\textbf{Conserved charges, angular amplitudes and some useful finite integral.}
As was discussed in the text, the background possess  four isometries  and string dynamics is invariant under shifts by arbitrary constants along to the directions $T,\, V,\,\psi $ and $\phi \equiv \phi_1$.
Taking into account the results of our considerations so far the momentum densities associated to these isometries can be written as
\begin{align}
&\frac{\alpha}{T}\,\Pi_T \,= \,-\,G_{TT}\,\kappa -G_{TV}\,(\omega_0 +\beta\,v')+\alpha\, B_{T \phi}\,\Phi' +\alpha\,B_{T\psi}\,\Psi' \,, \\
&\frac{\alpha}{T}\,\Pi_V\,=\,G_{TV}\,\kappa \,,\\
&\frac{\alpha}{T}\,\Pi_{\phi}\,=\, G_{\phi\phi} \,(\omega_1 +\beta\,\Phi')+ G_{\phi \psi} \,(\omega_3+\beta\,\Psi') \,,\\
&\frac{\alpha}{T}\,\Pi_{\psi}\,=\,  G_{\phi\psi} \,(\omega_1+\beta\,\Phi') +G_{\psi\psi}\,(\omega_3 +\beta\,\Psi') \,.
 \end{align}
where
\begin{align}
& v'(\xi) = -\,\frac{b\alpha\mu\omega_1}{(\alpha^2-\beta^2)} \,u\,,\label{eq-v u} \\
& \Phi'(\xi) = \frac{1}{(\alpha^2-\beta^2)} \left[\frac{2A}{b} \,\frac{1}{(1-u)} +\beta\omega_1 \right] , \label{eq-Phi u} \\
& \Psi'(\xi) = \frac{1}{(\alpha^2-\beta^2)} \left[ \frac{2A}{b} \,\frac{1}{(1-u)} + \frac{4(1-b)A}{b^2} - \frac{2\alpha\mu\kappa}{b Z^2}+\beta\omega_3 \right]. \label{eq-Psi u}
\end{align}
Actually, in order to find the dispersion relations we will need the explicit expression of the four momenta in terms of the variable $u$
\begin{align}
&\frac{\alpha}{T}\,\Pi_T\,=\, |G_{TT}|\kappa -\frac{\omega_0}{Z^2} +\frac{1}{(\alpha^2-\beta^2)} \frac{b\alpha\mu}{2Z^2}\left[ \frac{4}{b^2}\left( A - \frac{b\alpha\mu}{2Z^2}\kappa \right) +\beta(\omega_1 +\omega_3) \right] ,\label{moment T} \\
&\frac{\alpha}{T}\,\Pi_V\,= \,\frac{\kappa}{Z^2}\,, \label{moment V}\\
&\frac{\alpha}{T}\,\Pi_{\psi}\,= \,\frac{1}{(\alpha^2-\beta^2)}\left\lbrace \frac{b^2}{4}\left[ \beta\frac{4}{b^2} \left( A -\frac{b\alpha \mu}{2Z^2}\kappa \right) +\alpha^2\,(\omega_1+\omega_3) \right] -\frac{\alpha^2 b^2\omega_1}{2}\,u \right\rbrace, \label{moment psi}
\end{align}
\begin{multline}\label{moment phi}
\frac{\alpha}{T}\,\Pi_{\phi}\,=\, \frac{1}{(\alpha^2-\beta^2)}\left\lbrace \frac{b^2}{4}\left[ \beta\frac{4}{b^2} \left( A -\frac{b\alpha \mu}{2Z^2}\kappa \right) +\alpha^2\,(\omega_1+\omega_3) \right] \right. \\
\left.  +\alpha^2 \left[    b(1-b)\omega_1 -b^2 \frac{\omega_3}{2} + \frac{\beta}{\alpha} \frac{b\mu}{Z^2}\kappa \right] u - \alpha^2(1-b)\,b\,\omega_1 \, u^2 \right\rbrace .
\end{multline}
These expressions have to be substituted in the corresponding charge integrals 
\begin{equation} \label{charge-int}
E=\!\int\limits_{-\infty}^{+\infty} \!d\xi\, \Pi_T, \quad J_V=\!\int\limits_{-\infty}^{+\infty} \!d\xi\, \Pi_V,\quad J_\phi=\!\int\limits_{-\infty}^{+\infty} \!d\xi\, \Pi_\phi,   \quad J_\psi= \!\int\limits_{-\infty}^{+\infty} \!d\xi \,\Pi_\psi\,.
\end{equation}


As usually happens in this kind of considerations, some of the expression in \eqref{charge-int} are divergent. Indeed, the first two integrals are trivially divergent $\Pi_T$ and $\Pi_V$ being constants. Nevertheless, they may enter dispersion relations in certain combinations.
In order to find the dispersion relations for the giant magnon and single spike  string configurations we have to construct and compute the angular amplitudes along the directions of $\phi$ and $\psi$. To do that we have to integrate equations \eqref{eq-Phi u} and \eqref{eq-Psi u} over $\xi\in(-\infty, +\infty)$ on the solution \eqref{solution}. Since, the integral of $1/(1-u)$ over $\xi\in (-\infty,+\infty)$ is divergent on the solution, both amplitudes $\Delta \phi$ and $\Delta \psi$ are divergent quantities and we have to take care of that. To this end it proves useful (see for instance \cite{Hofman:2006xt}, \cite{Benvenuti:2008bd} and \cite{Georgiou:2017pvi}) to introduce the combination
\begin{equation}\label{delta}
\Delta\,\equiv \,\frac{\Delta \phi +\Delta \psi}{2}\,=\int\limits_{-\infty}^{+\infty} \!d\xi \left(\frac{\Phi'+\Psi'}{2} \right). 
\end{equation}
This amplitude $\Delta$ remains finite for the giant magnon case and it diverges for the single spike string configuration. In both cases, it has the form
\begin{equation}\label{delta u}
\Delta\,=\,\frac{1}{(\alpha^2-\beta^2)}\int\limits_{-\infty}^{+\infty} \!d\xi \left\lbrace \frac{1}{2}\left[ \frac{4}{b^2} \left(A - \frac{b\alpha \mu}{2Z^2}\kappa \right) + \beta(\omega_1+\omega_3) \right] +\frac{2A}{b} \left( \frac{u}{1-u}\right)  \right\rbrace .
\end{equation}

>From the explicit expressions for the momentum densities and amplitude $\Delta$ it is clear that the integrals we will need to calculate actually three.
Since the function $u(\xi)$  decrease when $\xi$ runs from $0$ to $+\infty$ and increase for $\xi$ running from $-\infty$ to $0$ (see \eqref{solution}), from \eqref{eq u roots} we can split the integration measure $d\xi$ in two ways
\begin{equation}
d\xi\,=\, -\, \frac{du}{u\,\sqrt{|a_{4}|\,(r_{+} -u)\,(u+|r_{-}|)}}\, , 
\end{equation}
when $\xi\in[0,+\infty]$ and the integration over $u$ has to be taken from $r_{+}$ to $0$. The integration over the other interval goes as 
\begin{equation}
d\xi\,=\, +\, \frac{du}{u\,\sqrt{|a_{4}|\,(r_{+} -u)\,(u+|r_{-}|)}}\, , 
\end{equation}
when $\xi\in[-\infty,0]$ and the integration over $u$ runs from $0$ to $r_{+}\,$.
Thus, in the symbolic form, we can write
\begin{equation}
\int\limits_{-\infty}^{+\infty} \!d\xi\,\cdots  \,=\,2\int\limits_{0}^{r_{+}} \frac{du}{u\,\sqrt{|a_{4}|\,(r_{+} -u)\,(u+|r_{-}|)}}\,\cdots.
\end{equation}
The three integrals we will need for coputation of the dispersion relations are 
\begin{align}
&Int_1\equiv\int\limits_{-\infty}^{+\infty} \!d\xi \,u\,=\,\frac{4}{\sqrt{|a_{4}|}}\,\arctan \sqrt{\frac{r_{+}}{|r_{-}|}}\,,\label{int u}\\
&Int_2\equiv\int\limits_{-\infty}^{+\infty} \!d\xi \,u^2\,=\,2\frac{(r_{+}-|r_{-}|)}{\sqrt{|a_{4}|}}\,\arctan \sqrt{\frac{r_{+}}{|r_{-}|}}+ 2\,\sqrt{\frac{r_{+}\,|r_{-}|}{|a_4|}}\,,\label{int u^2}\\
&Int_3\equiv\int\limits_{-\infty}^{+\infty} \!d\xi \,\frac{u}{1-u}\,=\,\frac{4}{\sqrt{|a_{4}|\,(1-r_{+} )\,(1+|r_{-}|)}}\,\arctan \sqrt{\frac{r_{+}(1+|r_{-}|)}{|r_{-}|(1-r_{+})}}\,.\label{int u/(1-u)}
\end{align}
In what follows we will need also the relation between the constant $A$ and roots $r_{\pm}\,$ \eqref{A^2 roots}

The expression for $\Delta$ needs some care. Explicitly, we have 
\begin{equation}\label{delta A}
\Delta= \frac{1}{(\alpha^2-\beta^2)}\left\lbrace \frac{1}{2}\left[ \frac{4}{b^2}\left(A - \frac{b\alpha \mu}{2Z^2}\kappa \right) +\beta(\omega_1+\omega_3) \right] \right\rbrace \!\int\limits_{-\infty}^{+\infty} \!\!d\xi
\,+\, 2\arctan \sqrt{\frac{r_{+}(1+|r_{-}|)}{|r_{-}|(1-r_{+})}}\,.
\end{equation}

To find the dispersion relations we will need some convenient parametrizations.
%
%
The clues come, for instance from \cite{Benvenuti:2008bd}, and indeed, to avoid the cumbersome dependence on the parameter we need conveniently defined quantities. To this end we define
\begin{align}
D\,\equiv\,\frac{r_+ - |r_-|}{r_+ +|r_-|}\,&\equiv\,-\cos\delta\,,\label{parametrization D}\\
C+D\,\equiv\,\frac{2\,r_+\,|r_-|}{(r_+ +|r_-|)}\,+\,\frac{r_+ -|r_-|}{r_+ +|r_-|}\,&\equiv\,-\cos\gamma\,,\,\,\,\, \delta\leq\gamma\in[0,\pi]\,,\label{parametrization C+D}
\end{align}
i.e. the form of solution \eqref{solution cosh} can be written as
\begin{equation}\label{solution cosh-a}
u(\xi)\,=\,\frac{C}{\cosh\left( \sqrt{a_2} \,\,\xi\right) -\, D}\,, \qquad a_2 >0\,, \qquad |C+D| <1\,.
\end{equation}
It is useful also to have at hand the realations
\begin{align}
\frac{r_{+}}{|r_{-}|}\,=\,\frac{1+D}{1-D}\,=\,\frac{1-\cos\delta}{1+\cos\delta}\,=\,\tan^2 \frac{\delta}{2}\,, \label{tg^2 delta}\\
\frac{r_{+}(1+|r_{-}|)}{|r_{-}|(1-r_{+})}\,=\,\frac{1+(C+D)}{1-(C+D)}\,=\,\frac{1-\cos\gamma}{1+\cos\gamma}\,=\,\tan^2 \frac{\gamma}{2}\,,\label{tg^2 gamma}
\end{align}
and 
\begin{equation}\label{r+r-}
r_+\,|r_-|\,=\,\frac{C^2}{1-D^2}\,=\,\frac{a_2}{|a_4|}\,.
\end{equation}
The important ratio $a_2/|a_4|$ can be written entirely in term of $\gamma$ and $\delta$  
\begin{equation}\label{cos delta -cos gamma}
\frac{\cos\delta \,-\, \cos\gamma}{\sin\delta}\,=\, \sqrt{\frac{a_2}{|a_4|}}\,.
\end{equation}

The expressions for the integrals \eqref{int u} and \eqref{int u^2} drastically simplify
\begin{align}
&Int_1\,=\,\frac{2}{\sqrt{|a_{4}|}}\,\delta\,, \label{Int_1}\\
&Int_2\,=\,\frac{a_3}{|a_4|\,\sqrt{|a_{4}|}}\,\delta + 2\frac{\sqrt{a_2}}{|a_4|}\,.\label{Int_2}
\end{align}
Moreover,
\begin{equation}\label{Int_2_1}
Int_2\,=\,\frac{a_3}{2\,|a_4|}\,Int_1 +2\,\frac{\sqrt{a_2}}{|a_4|}\,.
\end{equation}
The expression for $\Delta$ is
\begin{equation}\label{Delta gamma}
\Delta\,=\,\frac{1}{(\alpha^2-\beta^2)}\left\lbrace \frac{1}{2}\left[ \frac{4}{b^2}\left(A -\frac{b\alpha \mu}{2Z^2}\kappa \right) +\beta(\omega_1+\omega_3) \right] \right\rbrace \!\int\limits_{-\infty}^{+\infty} \!d\xi + \gamma \,.
\end{equation}

After these preparations, one can compute the general form of charges $J_\psi$ and $J_\phi$
\begin{equation}\label{charge psi}
\alpha\, \frac{J_\psi}{T} = \frac{1}{(\alpha^2-\beta^2)} \left\lbrace \frac{b^2}{4}\left[ \beta\frac{4}{b^2} \left(A - \frac{b\alpha\mu}{2Z^2}\kappa \right) + \alpha^2 \,(\omega_1+\omega_3) \right] \right\rbrace \!\int\limits_{-\infty}^{+\infty}\!d\xi \,-\, \frac{\alpha b^2}{2\,\sqrt{1-b}}\,\delta \,,
\end{equation}
\begin{equation}\label{charge phi}
\alpha\, \frac{J_\phi}{T} = \frac{1}{(\alpha^2-\beta^2)}\left\lbrace \frac{b^2}{4} \left[ \beta\frac{4}{b^2} \left(A -\frac{b\alpha\mu}{2Z^2}\kappa \right) + \alpha^2\,(\omega_1+\omega_3) \right] \right\rbrace \!\int\limits_{-\infty}^{+\infty} \!d\xi \,-\,\alpha b\sqrt{1-b} \sqrt{\frac{a_2}{|a_4|}}\,.
\end{equation}
Making use of \eqref{cos delta -cos gamma} one can rewrite the  charge $J_{\phi}$ as
\begin{multline}\label{charge phi delta gama}
\alpha\, \frac{J_{\phi}}{T} = \frac{1}{(\alpha^2-\beta^2)}\left\lbrace \frac{b^2}{4}\left[ \beta\frac{4}{b^2} \left(A -\frac{b\alpha \mu}{2Z^2}\kappa \right) + \alpha^2\,(\omega_1 +\omega_3) \right] \right\rbrace \!\int\limits_{-\infty}^{+\infty} \!d\xi \\
-\,\alpha b\sqrt{1-b} \left( \frac{\cos\delta -\cos\gamma}{\sin\delta} \right).
\end{multline}
In subsequent considerations, it should be taken into account that the quantity $\alpha^2-\beta^2$ is positive for the magnons and negative for the spikes. For the two cases it is useful to introduce the notation $\eta^2$ with value $\eta^2 \equiv \frac{\beta^2}{\alpha^2} $ for the magnons and $\eta^2\equiv \frac{\alpha^2}{\beta^2}$ for the spikes.

Now we are ready to obtain the dispersion relations for the two string solutions - giant magnon and spiky string configurations.

%
%
\subsection{Giant magnons}
In this subsection we calculate the charges and obtain the dispersion relations for the case of giant magnon strings. The boundary conditions ensuring giant magnon configuration of the string have been given in \eqref{magnon}, \eqref{A magnon}, \eqref{Z magnon} and \eqref{kappa magnon}. What remains to do is to insert them into the integrals for charges and obtain the dispersion relations. Substituting the relations from boundary conditions into the charge integrals one finds
\begin{align}
&\alpha\,\frac{E}{T}\,=\, \kappa\,\int\limits_{-\infty}^{+\infty} d \xi\,=\, \sqrt{\frac{\omega_0^2}{\mu^2} + \frac{b^2}{4}\, (\omega_1 + \omega_3)^2}\,\int\limits_{-\infty}^{+\infty} d\xi\,, \label{E mag} \\
&\alpha \, \frac{J_V}{T}\,=\,\frac{\omega_0}{\mu^2}\,\int\limits_{-\infty}^{+\infty} d \xi\,, \label{J V mag}\\
&\alpha\, \frac{J_{\psi}}{T}\,=\, \frac{b^2}{4}\, (\omega_1 + \omega_3)\,\int\limits_{-\infty}^{+\infty} d \xi\,  -\,\alpha\,\frac{b^2}{2\,\sqrt{1-b}}\,\delta \,, \label{J psi mag}\\
&\alpha\, \frac{J_{\phi}}{T} \,=\, \frac{b^2}{4}\, (\omega_1 + \omega_3)\,\int\limits_{-\infty}^{+\infty} d \xi\, -\,\alpha b\sqrt{1-b}  \left( \frac{\cos\delta \,-\, \cos\gamma}{\sin\delta} \right)  .\label{J phi mag}
\end{align}
The expression for $\Delta$ turns out to be finite
\begin{equation}\label{delta mag}
\Delta\,=\, \gamma  \,.
\end{equation}
However, due to constant terms in the integrands  all four  charges are divergent. Nevertheless, one can combine them into nice dispersion relations.

First, we observe that the divergence presented in all the charges cane be cast in the form
\begin{equation}\label{E-JV}
\frac{b}{2}\,(\omega_1 + \omega_3)\,\int\limits_{-\infty}^{+\infty} d \xi \,=\,\alpha\, \sqrt{\left( \frac{E}{T}\right) ^2 -\mu^2 \, \left(\frac{J_V}{T} \right)^2 }.
\end{equation}
Next step is, using \eqref{E-JV} to identify the finite combinations of charges. These are
\begin{align}
&\frac{J_{\psi}}{T} - \frac{b}{2} \sqrt{\left( \frac{E}{T}\right) ^2 -\mu^2 \, \left(\frac{J_V}{T} \right)^2 }\,,\qquad \frac{J_{\phi}}{T} - \frac{b}{2} \sqrt{\left( \frac{E}{T}\right) ^2 -\mu^2 \, \left(\frac{J_V}{T} \right)^2 }\,, \label{comb1-finite}\\
&\frac{J_{\psi}}{T}+\frac{J_{\phi}}{T} - b\,\sqrt{\left( \frac{E}{T}\right) ^2 -\mu^2 \, \left(\frac{J_V}{T} \right)^2 }\, , \qquad  \frac{J_{\psi}}{T}-\frac{J_{\phi}}{T}\,,\label{comb2-finite}
\end{align}
 but one should emphasize that only two of them are independent.
The first expression in \eqref{comb1-finite} is calculated to be
\begin{equation}\label{comb psi}
\frac{b}{2} \sqrt{\left( \frac{E}{T}\right) ^2 -\mu^2 \, \left(\frac{J_V}{T} \right)^2 }\,-\,\frac{J_{\psi}}{T} \,=\, \frac{b^2}{2\,\sqrt{1-b}}\,\delta\,.
\end{equation}
Subtracting the last expression in \eqref{comb2-finite} from the first one in \eqref{comb1-finite} gives
\begin{equation}\label{comb phi}
\frac{b}{2} \sqrt{\left( \frac{E}{T}\right) ^2 -\mu^2 \, \left(\frac{J_V}{T} \right)^2 }\,-\frac{J_{\phi}}{T} \,=\, b\sqrt{1-b} \left( \frac{\cos\delta \,-\, \cos\gamma}{\sin\delta} \right).
\end{equation}
Eliminating the parameter $\delta$ from the last two equations, \eqref{comb psi} and \eqref{comb phi}, and taking into account that $\,\Delta\,=\,\gamma\,$ we obtain the following dispersion relation for the giant magnon string configuration
\begin{multline}\label{dispers magnon}
 \frac{\cos\left\lbrace \sqrt{ \frac{(1-b)}{b^2} \left[ \left( \frac{E}{T}\right)^2 -\mu^2 \left(\frac{J_V}{T} \right)^2 \right] }    -\,\frac{2\,\sqrt{1-b}}{b^2} \left(\frac{J_{\psi}}{T}\right)  \right\rbrace \,-\,\cos\Delta\,}{\sin \left\lbrace \sqrt{ \frac{(1-b)}{b^2} \left[\left( \frac{E}{T}\right)^2 -\mu^2  \left(\frac{J_V}{T}\right)^2 \right] }    -\,\frac{2\,\sqrt{1-b}}{b^2} \left(\frac{J_{\psi}}{T}\right)  \right\rbrace }  \,=\\
 =\, \sqrt{\frac{1}{2(1-b)} \left[ \left( \frac{E}{T}\right)^2 - \mu^2 \left(\frac{J_V}{T} \right)^2 \right] } \,-\, \frac{1}{b\sqrt{1-b}} \left( \frac{J_{\phi}}{T}\right) .
\end{multline}
In the limit $\,\mu\,\rightarrow\,0\,$ we obtain the giant magnon dispersion relation on $\,AdS_5 \times T^{1,1}\,$ in \cite{Benvenuti:2008bd}. On the other hand, taking the limit $b\to 1$ ($Schr_5\times T^{1,1}\to Schr_5\times S^5$ limit) we find perfect agreement with the the result of \cite{Georgiou:2017pvi}.
%
%
\subsection{Single spikes}

Next issue is to obtain the dispersion relations for the single spike strings on $Schr_5\times T^{1,1}$ background. In the previous Section we determined the boundary conditions corresponding to spike string configurations, \eqref{spike}, \eqref{A spike}, \eqref{kappa spike} and \eqref{Z spike}.
The charges under these conditions have the form
\begin{align}
&\alpha\,\frac{E}{T}\,=\, \kappa \,\int\limits_{-\infty}^{+\infty} d \xi\,=\, \frac{\omega_0}{\mu} \,\int\limits_{-\infty}^{+\infty} d \xi\,, \label{E spike} \\
&\alpha \,\mu\, \frac{J_V}{T}\,=\,\frac{\omega_0}{\mu}\,\int\limits_{-\infty}^{+\infty} d \xi  + \alpha\,\frac{b\,(\omega_1 + \omega_3)}{2\,\beta} \,\int\limits_{-\infty}^{+\infty} d \xi \,, \label{J V spike}\\
& \frac{J_{\psi}}{T}\,=\,  -\,\frac{b^2}{2\,\sqrt{1-b}}\,\delta \,, \label{J psi spike}\\
&\frac{J_{\phi}}{T} \,=\,  -\,b\sqrt{1-b} \left( \frac{\cos\delta \,-\, \cos\gamma}{\sin\delta} \right)  .\label{J phi spike}
\end{align}
Beside $E$ and $J_V$, in this case $\Delta$ is also divergent 
\begin{equation}\label{Delta spike}
\Delta\,=\, - \frac{(\omega_1+\omega_3)}{2\,\beta}\,\int\limits_{-\infty}^{+\infty} d \xi + \gamma \,.
\end{equation}
However, combining all divergent expressions, \eqref{E spike}, \eqref{J V spike} and  \eqref{Delta spike} to cancel common divergences one can construct a finite combination 
\begin{equation}\label{gamma delta}
\gamma\,=\, \frac{1}{b}\,\left[   \mu \,\left(\frac{J_V}{T} \right) - \left(\frac{E}{T} \right) \right] +\Delta\,.
\end{equation}
The dispersion relations can be extracted from \eqref{J phi spike}. We have just to use \eqref{gamma delta} and \eqref{J psi spike} to eliminate $\gamma$ and $\delta$ parameters. The final expression for the dispersion relation of single spike string solutions is obtained to be
\begin{equation}\label{disp spike}
\frac{\cos\left[  \frac{2\,\sqrt{1-b}}{b^2} \, \left( \frac{J_{\psi}}{T} \right) \right] - \cos\left\lbrace  \frac{1}{b}\,\left[  \left(\frac{E}{T} \right) -  \mu \,\left(\frac{J_V}{T} \right)  \right] -\Delta \right\rbrace  }{\sin \left[  \frac{2\,\sqrt{1-b}}{b^2} \, \left( \frac{J_{\psi}}{T} \right) \right]}\,=\, \frac{1}{b\,\sqrt{1-b}} \left( \frac{J_{\phi}}{T} \right).
\end{equation}

This completes the derivation of the dispersion relations of giant magnon and single spike solutions of strings on $Schr_5\times T^{1,1}$ background.


\section{Concluding remarks}

In this Section we summarize our results and give some future directions.

The focus of our investigation in this paper were giant magnons and single spike string solutions in $Schr_5\times T^{1,1}$ background. These problems are important because their field theory duals are strongly coupled non-relativistic CFTs. The later are supposed to be dipole theories which seems to be far from completely understood.  Making use of finite combinations of conserved charges we were able to find the dispersion relations for these classes of string solutions in the above background. The main results are dispersion relations given in \eqref{dispers magnon} for giant magnons and \eqref{disp spike} for single spike strings, The natural way to go beyond these studies is to speculate what the field theory operators corresponding to the dispersion relations we obtained here would be. Certain clues come for instance, from \cite{Guica:2017jmq} or \cite{Georgiou:2017pvi}.

To make link to other studies on the subject we can consider several limiting cases, which was actually the reason to keep parameter $b$ in the formulas ($b=2/3$ for conifold). First of all, we would like to compare our results with those of the case of $Schr_5\times S^4$  \cite{Georgiou:2017pvi,Ahn:2017bio}. Looking at our line element it is easy to see that in the limit $b\to 1$ the spherical case is recovered. Carrying out carefully the limit in \eqref{dispers magnon} one finds
\begin{multline}\label{dispers magnon-limit}
\frac{\cos\left\lbrace \sqrt{ \frac{(1-b)}{b^2} \left[ \left( \frac{E}{T}\right)^2 -\mu^2 \left(\frac{J_V}{T} \right)^2 \right] }    -\,\frac{2\,\sqrt{1-b}}{b^2} \left(\frac{J_{\psi}}{T}\right)  \right\rbrace \,-\,\cos\Delta\,}{\sin \left\lbrace \sqrt{ \frac{(1-b)}{b^2} \left[\left( \frac{E}{T}\right)^2 -\mu^2  \left(\frac{J_V}{T}\right)^2 \right] }    -\,\frac{2\,\sqrt{1-b}}{b^2} \left(\frac{J_{\psi}}{T}\right)  \right\rbrace }  \,=\\
=\, \sqrt{\frac{1}{2(1-b)} \left[ \left( \frac{E}{T}\right)^2 - \mu^2 \left(\frac{J_V}{T} \right)^2 \right] } \,-\, \frac{1}{b\sqrt{1-b}} \left( \frac{J_{\phi}}{T}\right) 
\\     \qquad
\underset{b\to 1}{\Longrightarrow}  \qquad
 \left(\sqrt{\left( \frac{E}{T}\right)^2 -\mu^2 \left(\frac{J_V}{T} \right)^2} -\left(\frac{J_1}{T}\right) \right)^2-\left(\frac{J_2}{T}\right)^2 =
 4\sin^2\frac{\Delta}{2},
\end{multline}
where $J_1=J_\psi+J_\phi$ and $J_2=J_\psi-J_\phi$. As we see, the cumbersome transcendental relation greatly simplifies and we find perfect agreement with the results in \cite{Georgiou:2017pvi}. Analogous limit shows consistency for single spike case as well.

Another limit is the case of $\mu\to 0$ where the standard relations for conifold should be reproduced \cite{Benvenuti:2008bd}. This limit actually is easier and the agreement with confold case is obvious
\eq{
\eqref{dispers magnon}\quad \underset{\mu\to 0}{\Longrightarrow} \qquad
\frac{\cos\left\lbrace \frac{\sqrt{1-b}}{b}\left[ \frac{E}{T} - \frac{2}{b}\left(\frac{J_{\psi}}{T}\right)  \right]  \right\rbrace \,-\,\cos\Delta\,}{\sin \left\lbrace  \frac{\sqrt{1-b}}{b}\left[ \frac{E}{T} - \frac{2}{b}\left(\frac{J_{\psi}}{T}\right)  \right]  \right\rbrace }= \frac{ \frac{E}{T} - \frac{2}{b}\left(\frac{J_{\psi}}{T}\right)  }{2\sqrt{1-b}}
}

There are two more limits which could be taken - of point-like string (BMN, \cite{Berenstein:2002jq}) and folded string (GKP, \cite{Gubser:2002tv}). Let us briefly comment on the dispersion relation \eqref{dispers magnon} in these two limits.

The MBN limit the following quantities are small
\ml{
 \sqrt{ (1-b) \left[ \left( \frac{E}{T}\right)^2 -\mu^2 \left(\frac{J_V}{T} \right)^2 \right] }    -\,\frac{2\,\sqrt{1-b}}{b} \left(\frac{J_{\psi}}{T}\right) \: \sim  \\
 \: \sim \: \Delta \sim  \sqrt{ (1-b) \left[ \left( \frac{E}{T}\right)^2 -\mu^2 \left(\frac{J_V}{T} \right)^2 \right] } -\,\frac{2\,\sqrt{1-b}}{b} \left(\frac{J_{\phi}}{T}\right)  \:\to\:0.
}
Expanding the arguments of trigonometric functions one finds
\eq{
\left( \frac{E}{T} - \frac{2-b}{b}\left(\frac{J_{\psi}}{T}\right)- \frac{J_{\phi}}{T} \right)^2 = (J_\psi -J_\phi)^2 +b^2\Delta^2.
}
For conifold case actually we shoould set $b=2/3$
\eq{
\left( \frac{E}{T} - 2\left(\frac{J_{\psi}}{T}\right)- \frac{J_{\phi}}{T} \right)^2 = (J_\psi -J_\phi)^2 + \frac{4}{9}\Delta^2.
}
The left hand side of the above equation has been used in \cite{Itzhaki:2002kh,Gomis:2002km,PandoZayas:2002dso} to classify the states in the pp-wave limit with identification $H=E-2J_\psi-J_\phi$. To get exact agreement in the spherical limit however, one has to rescale $\Delta$. 
One can compare our result also with BMN strings considered in \cite{Guica:2017jmq} or \cite{Georgiou:2017pvi}.  For the first case one has to identify $J_V/T$ with $M$, set $J_\phi=0$ and take $E=\sqrt{\lambda}\kappa$. For the second case we just have first to take the limit $b\to 1$ ans since this limits is consistent with the results of \cite{Georgiou:2017pvi} further considerations also agree. 

It is interesting to mention the GKP regime which corresponds to $\eta\to0$. In this case one has to go back to the expression and carefully take the limit applied to the coefficients of the polynomial  \eqref{EqU} and then the value of $\Delta$ through \eqref{parametrization C+D} and \eqref{delta mag}. Then, one finds
\begin{multline}\label{dispers magnon-gkp}
 \sqrt{\frac{1}{2(1-b)} \left[ \left( \frac{E}{T}\right)^2 - \mu^2 \left(\frac{J_V}{T} \right)^2 \right] } \,-\, \frac{1}{b\sqrt{1-b}} \left( \frac{J_{\phi}}{T}\right) \\
= \cot\left\lbrace \sqrt{ \frac{(1-b)}{b^2} \left[ \left( \frac{E}{T}\right)^2 -\mu^2 \left(\frac{J_V}{T} \right)^2 \right] }    -\,\frac{2\,\sqrt{1-b}}{b^2} \left(\frac{J_{\psi}}{T}\right)  \right\rbrace.
\end{multline}
This is our expression for GKP limit. Note that, while in the point-like string regime the dispersion relation becomes quadratic, in the folded string limit it remains transcendental. 
Detailed study of these issues we leave for another paper.

Let us list some future directions.
Certainly the limits discussed above deserve further study especially with identifying the corresponding field operators. Actually this is the next step we intend to do. 

To complete the analysis we plan to investigate also pulsating strings in $Schr_5\times T^{1,1}$ background.
It would be interesting to study certain correlation functions and compare the results with other cases. Another issue is to study also finite correction  as it has been done in relativistic cases.

It would be interesting to study problems related to quantum information metric, complexity etc and we hope to return to these issues in the near future.

\paragraph{Acknowledgements}\ \\
R. R. is grateful to Kostya Zarembo for discussions on various issues of holography in Schr\"odinger backgrounds. T. V. and M. R are grateful to Prof. G.~Djordjevic for the warm hospitality at the University of Ni\v{s}, where some of these results have been presented. The work is partially supported by the Program “JINR– Bulgaria”  at  Bulgarian Nuclear Regulatory Agency. This work was supported in part by BNSF Grant DN-18/1 and H-28/5, as well as SU Grants 80-10-62/2020 and 80-10-68/2020.


\begin{appendix}
\label{appA}

\section{Lightning Review of Schr\"odinger spaces and general deformations }

\subsection{A remark on deformations}

One of the most general deformations preserving integrability  is the so-called Drinfe’ld-Reshetikhin (DR) twist, particular case of which TsT transformation appears to be. Let us briefly mention the main features of DR twists following mainly \cite{Guica:2017jmq}.
		
\indent The DR twist of the scattering matrix $\mathbb{S}$ is realized as
\eq{
\mathbb{S}\:\rightarrow\: \tilde{\mathbb{S}}=\mathbb{F}\,\mathbb{S}\,\mathbb{F},\qquad \mathbb{F}=e^{\frac{i}{2}\sum_{i,j}\gamma_{ij}(H_i\otimes H_j-H_j\otimes H_i)},
}
where $H_i$ are Cartan elements of the isometry group and $\gamma_{ij}$ is a constant antisymmetric matrix. In the cases above, instead of coefficient times Cartan element, we have the corresponding charges. An important oint to stress on is that the element of the Cartan matrix can be replaced by commuting (super) charges and the twist still preserves integrability.

The construct a DR twist one can be use the R-matrix, which acts on the tensor product of two vector spaces:
\eq{
R_{ab}(u): \quad V_a\otimes V_b\:\longrightarrow	\: V_a\otimes V_b,
}
and satisfy the Yang-Baxter (YB) equation
\eq{
R_{ab}(u-v)R_{ac}(u)R_{bc}(v)=R_{bc}(v)R_{ac}(u)R_{ab}(u-v).
}
Then the statement is that the Drinfe'ld twist is the most general linear transformation preserving integrability and has the form
\eq{
			R_{ab}(u)\:\longrightarrow\: \tilde{R}_{ab}(u)= F_{ab}R_{ab}(u)F_{ab}, 
		}
The constant matrix $F_{ab}$ satisfies the following conditions:
\begin{itemize}
\item $F_{ab}$ is a constant solution of YB equations:
\eq{F_{ab}F_{ac}F_{bc}=F_{bc}F_{ac}F_{ab}.}
\item Obeys associativity condition: 
\eq{
R_{ab}(u)F_{ca}F_{cb}=F_{cb}F_{ca}R_{ab}(u).}
\item To preserve regularity of the R-matrix, the twist should satisfy an unitarity condition of the form:
\eq{
F_{ab}F_{ba}=1.}
\end{itemize}

In practice the above construction work as follows. Lets have a set of commuting in $V_a$ charges $Q^i_a$, $[Q^i_a,Q^j_a]=0$. Thus, the condition 
\eq{
e^{i\omega_k Q^k_a}e^{i\omega_l Q^l_b}R_{ab}(u)= R_{ab}(u)e^{i\omega_k Q^k_a}e^{i\omega_l Q^l_b}\qquad \text{for each}\: k,l
}
means that $e^{i\omega_k Q^k_a}$ is a non-degenerate linear transformation on $V_a$ and is a symmetry of $R_{ab}$.
Defining $K_a=e^{i\omega_i Q^i_a}$,  it is a simple exercise to check that  $F_{ab}=K_a K^{-1}_b$ satisfies YB equation.

In summary, one can construct the Drinfe'ld twist operator as (summation over $i,j$ is understood):
\eq{
F_{ab}=e^{\frac{i}{2}\gamma_{ij}Q^i_aQ^j_b},
}
where $\gamma_{ij}=-\gamma_{ji}$.

Let us turn to TsT transformations and assume that we have a background with associated brane system. To implement TsT transformation at least two isometry directions are needed, say $(\phi_1,\phi_2)$. The transformation then consists of a T-duality along $\phi_1$, followed by a shift $\phi_2 \rightarrow \phi_2 + \gamma\phi_1$ in the T-dual background and T-duality back along $\phi_1$. Depending on where the ismetries lies, one can distinguish three cases:
\begin{enumerate}
	\item The first case is when the two isometries involved in the TsT-transformation along along the brane. In this case the product of the fields in dual gauge theory becomes:
	\begin{eqnarray}
	(f\ast g)(x)& =& e^{-i\pi\gamma\left(
		\frac{\partial}{\partial x^1}\frac{\partial}{\partial y^2}-\frac{\partial}{\partial x^2}
		\frac{\partial}{\partial y^1}\right)}
	f(x)g(y)_{|x=y}\nonumber \\
	& = & f(x)g(x) - i\pi\gamma\left(\partial_{1}f(x)\partial_2g(x)-
	\partial_2f(x)\partial_1g(x)\right)+\cdots
	\end{eqnarray}
	Since gamma is constant, this is nothing but the Moyal product for a non-commutative two-torus. Obviously it is non-local and breaks Lorentz invariance and causality. Nevertheless, this picture, and its generalizations, offers interesting string realizations of non-commutative theories. 
	
	\item  To make the second class of deformations we assume that one global $U(1)$ isometry is along the D-brane, but the other one is transversal to the brane.  The deformed product of the fields in this case can be read off:
	\begin{equation}
	(f\ast g)(x)=e^{\pi\gamma\left(
		Q^g\frac{\partial}{\partial x}-Q^f\frac{\partial}{\partial y^1}\right)}=
	f(x+\pi\gamma Q^g)g(x-\pi\gamma Q^f),
	\end{equation}
	where $Q^i$ are the charges associated to the isometries. 
	This is called dipole deformation. As we can see it is clearly non-local in one direction, but still living on a commutative space-time.
	\item  In the last case both isometries transverse to the D-brane. The product in th dual gauge theory but  the two charges does not act as derivatives anymore
	\begin{equation}
	(f\ast g)(x)=e^{i\pi\gamma(Q^f_1Q^g_2-Q^f_2Q^g_1)}fg. \label{2.6}
	\end{equation}
	The product deformation yields an ordinary commutative and local theory,
	since the only contribution to the deformed product are some phases. The superconformal gauge theories, arising from the product \eqref{2.6}, are classified by Leigh and Strassler and are called $\beta$-deformed.
	As we discussed above,  the deformation reduces the amount of supersymmetry and TsT-transformations serves as supersymmetry breaking procedure as well.
\end{enumerate}

As it is clear from above, TsT transformation can be thought of as particular case of DR twist.


\subsection{$Schr_5\times S^5(T^{1,1})$ in Global coordinates} \label{appSchro}

The metric of 5D Schr\"odinger spacetime in local coordinates can be written as
\eq{
	ds^2=-\ell^2 \frac{{\hat\mu}^2(dx^+)^2}{z^4} + ds^2_{AdS_5},
	\label{schro-metric-a}
}
where the second part is the $AdS_5$ metric in  light-cone coordinates 
\begin{equation}
\label{A.1}
ds^2_{AdS_5}=\frac{\ell^2}{z^2}\left(2dx^+dx^-+d\vec{x}^2+dz\right).
\end{equation}
To obtain the metric in global coordinates we apply the following transformations
\begin{equation}\label{eq_A2}
x^+=\tan T,\quad x^-=V-\frac{1}{2}\left(Z^2+\vec{X}^2\right)\tan T,\quad z=\frac{Z}{\cos T},\quad \vec{x}=\frac{\vec{X}}{\cos T}.
\end{equation}
Thus 
\begin{align}
\label{A.3}
&dx^+=\frac{dT}{\cos^2T},\qquad dx^-=dV-\tan T\left(ZdZ+\vec{X}.d\vec{X}\right)-\frac{1}{2}\left(Z^2+\vec{X}^2\right)\frac{dT}{\cos^2T},\nonumber\\
&dz=\frac{dZ}{\cos T}+\frac{Z\sin TdT}{\cos^2T},\qquad d\vec{x}=\frac{1}{\cos T}\left(d\vec{X}+\vec{X}\tan TdT\right).
\end{align}
Substituting \eqref{A.3} into \eqref{A.1} we find the $AdS_5$ piece in global coordinates
\begin{equation}
ds^2_{AdS_5}=\frac{\ell^2}{Z^2}\left(2dTdV-(Z^2+\vec{X}^2)dT^2+d\vec{X}^2+dZ^2\right).
\end{equation}
For the rest, namely the first term in \eqref{schro-metric-a}, we just need to use the relation 
\begin{equation}
\frac{\hat{\mu}^2}{z^4}{dx^+}^2=\frac{\hat{\mu}^2}{Z^4}{dT}^2.
\end{equation}
Putting everything together we obtain the Schr\"odinger metric and the $B-$field in global coordinates 
\begin{equation}
\frac{ds^2_{Schr_5}}{\ell^2}=-\left(\frac{\hat{\mu}^2}{Z^4}+1 \right)dT^2+
\frac{2dT\,dV-\vec{X}^2dT^2+d\vec{X}^2+dZ^2}{Z^2},
\label{metric-schro-global}
\end{equation}
\begin{equation}
\alpha' B_{(2)}= \frac{\ell^2\hat{\mu}\, dT}{Z^2}\wedge (d\hat{\chi}+P).
\label{B-global-1}
\end{equation}

In the procedure outlined above the only element from transverse space $S^5$ participating the derivation is the isometry angle $\chi$. Therefore, to obtain the background $Schr_5\times T^{1,1}$ in global coordinates we need only that isometry angle from five-torus $T^{1,1}$.


\section{Collection of some formulae and details}\label{calculations}

In this appendix, we present some details of the derivation of the equations of motion.

Since the ansatz \eqref{ansatz} is linear in worldsheet time, the equations of motion for all non-trivial 2d fields become actually equations with respect to $\xi$ and all the constants $A_T,A_V,A_{\phi_1}, A_{\phi_2},$  and $A_{\psi}$ below are integration constants. They read off as follows\\

--- For T ---
\begin{multline}
\dfrac{\pa}{\pa \xi} \left\lbrace G_{TT}\left[ -\beta\,\dot{T} +(\alpha^2-\beta^2) \,{T'}\right] + G_{TV}\left[ -\beta\,\dot{V}+ (\alpha^2-\beta^2) \,V' \right]  \right.  \\
\left.  - \,\alpha \sum\limits_{i=1}^2 B_{T\phi_i} \dot{\phi_i} -\alpha \,B_{T\psi} \dot{\psi} \right\rbrace =0\, ,
\end{multline} 
or
\begin{multline}\label{v'-eq}
\left(1+ \frac{\mu^2}{Z^4}\right) \left[\beta\kappa -(\alpha^2-\beta^2)t' \,\right] -\frac{1}{Z^2} \left[\beta\omega_0 - (\alpha^2-\beta^2)v' \right] \\
-\frac{\alpha b\mu}{2Z^2} \left( \omega_3 -\sum\limits_{i=1}^2 \omega_i \,\cos\theta_i \right)= \, A_T.
\end{multline}

--- For V ---
\begin{equation}
\dfrac{\pa}{\pa \xi} \left\lbrace  G_{TV}\left[  -\beta\,\dot{T} +(\alpha^2-\beta^2) \,{T'}\right]\right\rbrace =0\,,
\end{equation}
or
\begin{equation}\label{t' Constant}
(\alpha^2-\beta^2) \,{t'}(\xi)=A_V Z^2 +\beta \kappa.
\end{equation}

Then, substituting \eqref{t' Constant} into \eqref{v'-eq}  we obtain the equation for $v'$
\begin{equation}\label{EOM-TV}
(\alpha^2-\beta^2)\,v'(\xi) = \frac{\alpha b\mu}{2} \left( \omega_3 -\sum\limits_{i=1}^2 \omega_i \,\cos\theta_i (\xi)\right) + A_T Z^2 +(Z^4+\mu^2)A_V+\beta\omega_0\,.
\end{equation}

--- For Z ---

\begin{multline}
\frac{\pa G_{TT}}{\pa Z} \left[   -\dot{T}^2 -2\beta\,\dot{T}T' +(\alpha^2-\beta^2) \,{T'}^2 \right] \\
+2\,\frac{\pa G_{TV}}{\pa Z} \left[ -\dot{T}\dot{V}-\beta\,(\dot{T}V'+\dot{V}T')+(\alpha^2-\beta^2 ) \, T'V' \right] \\
+ 2\alpha \sum\limits_{i=1}^2 \frac{\pa B_{T\phi_i}}{\pa Z} \left[\dot{T}{\phi_i}'-T'\dot{\phi_i} \right] +\, 2\alpha \,\frac{\pa B_{T\psi}}{\pa Z} \left[\dot{T}{\psi}'-T'\dot{\psi} \right]\,=\,0\,,
\end{multline}
or, explicitly 
\begin{multline}\label{EqZ}
\frac{\mu^2}{Z^2} \left[ \kappa^2 +2\beta\kappa\,t'-(\alpha^2-\beta^2) \,{t'}^2\right] - \left[\kappa\omega_0 + \beta\left( \kappa\,v'+\omega_0\, t'\right) - (\alpha^2-\beta^2) \,t'\,v'\, \right]  
\\
-\frac{\alpha b\mu}{2} \left[\, \omega_3-\sum\limits_{i=1}^2 \omega_i \,\cos\theta_i \right] t' - \frac{\alpha b\mu\kappa}{2} \left[ \,\sum\limits_{i=1}^2\, \Phi_i'\,\cos\theta_i -\Psi'\right] \,=0.
\end{multline}

--- For $\phi_k\,\qquad k=1,2$ ---

\begin{multline}
\dfrac{\pa}{\pa \xi}\left\lbrace  \sum\limits_{i=1}^2 G_{\phi_k \phi_i} \left[-\beta\,\dot{\phi_i} + (\alpha^2-\beta^2)\,{\phi_i}' \right] - G_{\phi_k \psi} \left[ \beta\,\dot{\psi} - (\alpha^2-\beta^2)\,{\psi}'\right] +\alpha \,B_{T\phi_k}\,\dot{T}  \right\rbrace =0\, ,
\end{multline}
or, as equations in $\xi$
\begin{multline}\label{EqPhi_k}
\frac{b^2}{4}\cos\theta_k \left\lbrace \beta \left( \omega_3 - \sum\limits_{i=1}^2 \omega_i \,\cos\theta_i \right) + (\alpha^2-\beta^2) \left( \sum\limits_{i=1}^2 \Phi_i' \cos\theta_i - \Psi'\right) \right\rbrace \\
-\frac{b}{4} \sin^2\theta_k \left[ \,\beta \omega_k - (\alpha^2-\beta^2) \,\Phi_k'\, \right] \,-\, \frac{\alpha b \mu\kappa}{2Z^2} \cos\theta_k \,=\, A_{\phi_k} \,, \qquad k=1,2\,.
\end{multline}

--- For $\psi$ ---

\begin{multline}
\dfrac{\pa}{\pa \xi} \left\lbrace \sum\limits_{i=1}^2 G_{\phi_i\psi} \left[   -\beta\,\dot{\phi_i} + (\alpha^2 - \beta^2)\,{\phi_i}^{\prime}\right] - G_{\psi\psi} \left[\,   \beta\,\dot{\psi} - (\alpha^2-\beta^2)\,{\psi}'\right] +\alpha \,B_{T\psi}\,\dot{T}  \right\rbrace =0\, ,
\end{multline}
or
\begin{equation}\label{EqPsi}
\beta \left( \omega_3 -\sum\limits_{i=1}^2 \omega_i \,\cos\theta_i \right) + (\alpha^2-\beta^2) \left( \sum\limits_{i=1}^2 {\Phi_i}' \cos\theta_i - \Psi'\right)
=\frac{4}{b^2} \!\left( \frac{\alpha b \mu\kappa}{2Z^2}  - A_{\psi} \right)  .
\end{equation}

Combining the EoM for $\psi$  \eqref{EqPsi} and EoM's for $\phi_k$ \eqref{EqPhi_k} we obtain the equations for $\Phi_k (\xi)$
\begin{equation}\label{EqPhi_k Independent}
(\alpha^2 - \beta^2)\,{\Phi_k}^{\prime}(\xi)\, =\,\frac{4}{b}\,\frac{(A_{\phi_k} +  A_{\psi}\,\cos{\theta_k})}{\sin^2 {\theta_k }} \,+\, \beta\,\omega_k  \,, \qquad k=1,2\,,
\end{equation}
while for $\Psi(\xi)$ it is
\begin{equation}\label{EqPsi Independent}
(\alpha^2-\beta^2)\,{\Psi}' (\xi)=\frac{4}{b} \sum\limits_{i=1}^2\dfrac{A_{\phi_i}\,\cos\theta_i +A_\psi}{\sin^2 \theta_i} +\dfrac{4(1-2b)}{b^2}\,A_\psi-\frac{2\alpha\mu\kappa}{bZ^2} +\beta\omega_3\,.
\end{equation}
Substituting these equations into the equation for Z \eqref{EqZ} we obtain a relation between the integration constants 
\begin{equation}\label{RelationZ-app}
A_V^2 \,Z^6 \,+\, A_V A_T \,Z^4 \,+\, \frac{\alpha\kappa}{b} \left(2\mu A_{\psi} - \alpha b\omega_0 \right)=0.
\end{equation}

Putting everything together, we obtain the following list of equations\\

---  $t^{\prime}$ ---
\begin{equation}\label{eq_t-app}
(\alpha^2 - \beta^2) \,{t^{\prime}}(\xi)=A_V Z^2 +\beta \kappa.
\end{equation}

--- $v^{\prime}$ ---
\begin{equation}\label{eq_v-app}
(\alpha^2 - \beta^2)\,v^{\prime}(\xi)\,= \frac{\alpha b\mu}{2} \left( \omega_3 - \sum\limits_{i=1}^2 \omega_i \,\cos\theta_i \right)
+ A_T Z^2  + (Z^4+\mu^2)A_V+\beta\omega_0\,.
\end{equation}

---  $\Phi_k^{\prime}$ ---
\begin{equation}\label{eq_Phi_k-app}
(\alpha^2 - \beta^2)\,{\Phi_k}^{\prime}(\xi)\, =\,\frac{4}{b}\,\frac{(A_{\phi_k} +  A_{\psi}\,\cos{\theta_k})}{\sin^2 {\theta_k }} \,+\, \beta\omega_k  \,, \qquad k=1,2\,.
\end{equation}

---  $\Psi^{\prime}$ ---
\begin{equation}\label{eq_Psi-app}
(\alpha^2-\beta^2)\,\Psi'(\xi)=\frac{4}{b}\sum\limits_{i=1}^2\dfrac{A_{\phi_i}\,\cos\theta_i +A_{\psi}}{\sin^2 {\theta_i}} +\dfrac{4(1-2b)}{b^2}\,A_{\psi}-\frac{2\alpha\mu\kappa}{b Z^2} +\beta\omega_3\,.
\end{equation}

---  $ Z $ ---
\begin{equation}\label{RelationZ}
A_V^2 \,Z^6 \,+\, A_V A_T \,Z^4 \,+\, \frac{\alpha\kappa}{b} \left(2\mu A_{\psi} - \alpha b\omega_0 \right)=0.
\end{equation}

--- For $\theta_k\,\qquad k=1,2$ ---
\begin{multline}
\frac{b}{4}(\alpha^2-\beta^2)\,{\theta_k}^{\prime\prime}(\xi) + \dfrac{\pa G_{\phi_k\phi_k}}{\pa \theta_k}\,\left[ \,\dot{\phi_k}^2 + 2\beta \,\dot{\phi_k}  {\phi_k}' -(\alpha^2-\beta^2)\,  {{\phi_k}'}^2 \right]\\
+\,\dfrac{\pa G_{\phi_1\phi_2}}{\pa \theta_k} \left[ \dot{\phi_1}\dot{\phi_2} + \beta \,(\dot{\phi_1}  {\phi_2}'+\dot{\phi_2}  {\phi_1}') -(\alpha^2-\beta^2)\,  {\phi_1}' {\phi_2}' \right] \\
+\,\dfrac{\pa G_{\phi_k\psi}}{\pa \theta_k} \left[ \dot{\phi_k}\dot{\psi} +\beta \,(\dot{\phi_k}  \psi'+\dot{\psi}{\phi_k}') - (\alpha^2-\beta^2) \,{\phi_k}' \psi' \right]\\
-\alpha\,\dfrac{\partial B_{T\phi_k}}{\partial \theta_k}\, \left[  \dot{T}{\phi_k}'-T^{\prime}\dot{\phi_k} \right] =0\, ,
\end{multline}
or
\begin{multline}
(\alpha^2 - \beta^2)\,{\theta_1}^{\prime\prime}(\xi) +(1-b)\cos\theta_1\sin\theta_1 \left[ \,\omega_1^2 +2\beta\omega_1 \Phi_1' -(\alpha^2-\beta^2)\,  {\Phi_1'}^2    \right] \\
-b\sin\theta_1\cos\theta_2 \left[ \,\omega_1\omega_2 +\beta \,(\omega_1\Phi_2'+\omega_2  \Phi_1') -(\alpha^2-\beta^2)\,  \Phi_1' \Phi_2'   \right] \\
+b\sin\theta_1 \left[ \,\omega_1\omega_3 +\beta \,(\omega_1 \Psi'+\omega_3  \Phi_1') -(\alpha^2-\beta^2)\,  \Phi_1' \Psi' \right] \\
-\frac{2\alpha\mu}{Z^2} \sin\theta_1 \left[  \kappa\Phi_1'-t'\omega_1 \right] =0\, ,
\end{multline}
\begin{multline}
(\alpha^2-\beta^2)\,{\theta_2}^{\prime\prime}(\xi) +(1-b)\cos\theta_2\sin\theta_2 \left[ \,\omega_2^2 + 2\beta\omega_2 \,\Phi_2' -(\alpha^2-\beta^2)\, {\Phi_2'}^2    \right]\\
-b\sin\theta_2\cos\theta_1 \left[ \,\omega_1\omega_2 +\beta \,(\omega_1  \Phi_2'+\omega_2  \Phi_1') -(\alpha^2-\beta^2)\,  \Phi_1'\Phi_2' \right] \\
+b\sin\theta_2 \left[ \,\omega_2\omega_3 +\beta \,(\omega_2  \Psi'+\omega_3  \Phi_2') -(\alpha^2-\beta^2)\,  \Phi_2'\Psi' \right] \\
-\frac{2\alpha\mu}{Z^2} \sin\theta_2 \left[  \kappa{\Phi_2}^{\prime}-t^{\prime}\omega_2 \right] =0\, .
\end{multline}
%

\end{appendix}


\end{document}